\documentclass[%
 reprint,
 amsmath,amssymb,
 aps,
 prd,
]{revtex4-1}
\usepackage[utf8]{inputenc}
\usepackage[colorlinks, pdfborder={0 0 0}]{hyperref}
\usepackage{xcolor}
\usepackage{graphicx}
\usepackage{amsmath,amssymb,amsbsy,amstext,amsthm}
\usepackage{mathtools}
\usepackage{amsfonts}
\usepackage{comment}
\usepackage{bm}
\usepackage{tensor}

\begin{document}

\title{Simulating gravitational waves passing through the spacetime of a black hole}

\author{Jian-hua He$^{1,2}$}
\thanks{Corresponding author:
\href{mailto:hejianhua@nju.edu.cn}{hejianhua@nju.edu.cn}}
\author{Zhenyu Wu$^{1,2}$}
\affiliation{$^1$School of Astronomy and Space Science, Nanjing University, Nanjing 210023, P. R. China}
\affiliation{$^2$Key Laboratory of Modern Astronomy and Astrophysics (Nanjing University), Ministry of
Education, Nanjing 210023, P. R. China}

\begin{abstract}
We investigate how GWs pass through the spacetime of a Schwarzschild black hole using time-domain numerical simulations. Our work is based on the perturbed 3+1 Einstein's equations up to the linear order. We show explicitly that our perturbation equations are covariant under infinitesimal coordinate transformations. Then we solve a symmetric second-order hyperbolic wave equation with a spatially varying wave speed. As the wave speed in our wave equation vanishes at the horizon, our formalism can naturally avoid boundary conditions at the horizon. Our formalism also does not contain coordinate singularities and, therefore, does not need regularity conditions. Then, based on our code, we simulate both finite and continuous initially plane-fronted wave trains passing through the Schwarzschild black hole. We find that for the finite wave train, the wave zone of GWs is wildly twisted by the black hole. While for the continuous wave train, unlike geometric optics, GWs can not be sheltered by the back hole. A strong beam and an interference pattern appear behind the black hole along the optical axis. Moreover, we find that the back-scattering due to the interaction between GWs and the background curvature is strongly dependent on the direction of the propagation of the trailing wavefront relative to the black hole.  Finally, for a realistic input waveform generated by binary black holes, we find that the lensed waveform in the merger and ringdown phases is much longer than that of the input waveform due to the effect of back-scattering.
\end{abstract}

\maketitle
\section{introduction}
“stellar-mass black holes” are formed from the deaths of high-mass ($> 8M_{\odot}$) stars, which are estimated as many as $10^8 \sim 10^9$ in our Milky Way~\cite{Agol:2001hb}. The majority of these black holes are expected to be in an isolated situation due to the disruption of their progenitor systems~\cite{Belczynski_2004} that exist mostly in binaries or in multiple systems~\cite{sana_2016}. However, despite their vast population, they are difficult to  detect, as unlike binary black holes, isolated black holes do not produce detectable emissions of their own. At the moment, the primary tools to detect them rely on either photometric microlensing~\cite{Bennett_2002,Mao} or astrometric microlensing (e.g.~\cite{2022arXiv220201903L}).

On the other hand, the discovery of gravitational waves ushered us into a new era of astronomy~\cite{Abbott:2016blz}. In the coming decades, ground- and space-based GW experiments will study the GW phenomena in unprecedented detail, such as the Einstein Telescope (ET)~\cite{Einstein_tele}, 40-km LIGO~\cite{LIGO40}, eLISA~\cite{2013arXiv1305.5720E}, DECIGO~\cite{Sato_2009}, and Pulsar Timing Arrays (PTA)~\cite{2010CQGra..27h4013H}. These experiments may provide a potential way beyond the conventional astronomical methods to detect isolated black holes.

On the theoretical aspect, since an isolated black hole does not emit detectable GWs itself, one possible way to detect it is when some external GWs pass through it and leave some detectable features in the waveforms. The key question then becomes how to accurately predict the waveforms of GWs when they pass through a black hole. 

In the weak field limit, such a problem has been studied in the previous work ~\cite{PhysRevD.34.1708,Deguchi,Schneider,Ruffa_1999,DePaolis:2002tw,Takahashi:2003ix,1999PThPS.133..137N,Suyama:2005mx,Christian:2018vsi,Zakharov_2002,Liao:2019aqq,Macquart:2004sh,PhysRevLett.80.1138,Dai:2018enj,PhysRevD.90.062003,Yoo:2013cia,Nambu:2019sqn}.
In these pioneer works (e.g.~\cite{Schneider}), a thin-lens model is assumed, in which the incident GWs are assumed to be far away from the optic axis and the impact parameter is much larger than the Schwarzschild radius of the lens mass $R_s$. The gravitational field of the black hole is weak in this case, and the deflection angle due to the black hole is small. In addition to the assumption of thin-lens, to address the wave effects of GWs, the previous work also applied {\it Kirchhoff's} diffraction theory to the lensing system. When the wavelength of GWs $\lambda$ is much smaller than the Schwarzschild radius $\lambda\ll R_s$, the diffraction integral is dominated by the stationary phase points. These points can be viewed as corresponding to distinct images. The approximation is known as the stationary phase approximation or the geometric optics approximation.

Based on the thin lens model and geometric optics approximation, most recently, a comprehensive analysis of lensing has been performed using the data from the first half of the third LIGO–Virgo observing run~\cite{LIGOScientific:2021izm}. However, no compelling evidence of lensing has been found.

In the opposite limit, when the wavelength of GWs is comparable to or much greater than the Schwarzschild radius $\lambda\ge R_s$, diffraction becomes significant. In this case, even in the weak field limit, the thin lens model and the geometric optics approximation break down for waves along the optic axis. GWs do not form caustics behind the lens at scales that are comparable to their wavelength. Instead, GWs form a strong beam along the optic axis~\cite{He:2021hhl}. 

In strong gravity, the situation is more complicated due to the complexity of the black hole spacetime. To fully address GWs passing through the spacetime of strong gravity,  we would like to investigate the wave property of GWs in the spacetime of a black hole first, which plays an important role here.

In flat spacetime, GWs are described by
\begin{equation}
\frac{\partial^2}{\partial t^2}h_{ij}-c^2\nabla^2 h_{ij} =c^2 Q_{ij}(t,x)  \,,\label{vaccumwave1}
\end{equation}
where $Q_{ij}(t,x)$ is the source mass and $c$ is the speed of light in vacuum. The wave equation has a fundamental solution $G$ which satisfies
\begin{equation}
\frac{\partial^2}{\partial t^2}G - c^2\nabla^2 G  =c^2\delta(t)\delta(x) \,.\nonumber
\end{equation}
For an outgoing wave, the explicit form of $G$ is known  as the {\it retarded Green's function} 
\begin{equation}
	G^{+}(x,t;x',t')=\frac{\delta(t'-[t-\frac{\left\|x-x'\right\|}{c}])}{4\pi\left\|x-x'\right\|}\,.\nonumber
\end{equation}
Unlike the Green's function in the 2D case (or even dimensions), the 3D (or odd dimensions) Green function has a notable feature, namely, the signal is ``sharp". A perturbation at a point $\vec{x}$ is visible at another point $\vec{x}'$ exactly at the time $t = |\vec{x} - \vec{x}'|/c$. Since the wave speed of GWs equals that of light rays, GWs travel exactly on the future direct light cone. Moreover, since the speed of GWs does not depend on frequency, there is no dispersion in a wavelet of GW signals. This leads to the strong Huygens' principle in the vacuum of flat spacetime~\cite{1965ArRMA..18..103G}: a finite GW wave signal has a definite wave zone with clear leading and trailing wavefronts. 

However, the strong Huygens' principle does not hold in strong gravity. A strict mathematical theorem states that in the four-dimensional spacetime the wave equation on an empty spacetime with a vanishing Ricci tensor satisfies Huygens' principle if and only if the spacetime is flat or is that of a plane wave \cite{1965ArRMA..18..103G,1990GReGr..22..843W,mclenaghan_1969}. As such, unlike in the case of flat spacetime, in strong gravity, besides the original signals, GWs can be scattered back due to the interaction between GWs and the background curvature~\cite{1992JMP....33..625S,1975weoc.book.....F}. As a result, a ``sharp" signal is no longer ``sharp" but can be dispersed, which generates a ``lasting" effect and leaves a tail of GW signals. This tail blurs the trailing wavefront. In this case, GWs propagate not only on the light cone but also inside it~\cite{1992JMP....33..625S,Poisson:2003nc}. 

Unlike the trailing wavefront, the leading wavefront represents the transfer of energy from one place to another, which is subject to the constraint of causality. Its propagation does not depend on the frequencies of GWs. This is because the leading wavefront is not necessarily continuous but can be with a sharp edge (e.g. square waves). In mathematics, the Fourier series, indeed, converge only in the sense of $L_2$ norm, which means that they converge up to functions with differences on a set of Lebesgue measure zero. They do not necessarily converge in the pointwise sense unless the function is continuously differentiable. Thus, the sum of the Fourier series does not converge near the point of discontinuity of a piecewise smooth function, which is known as the Gibbs phenomenon. Moreover, unlike in the 1D case, the wavefront in 3D space may have a complex geometric shape at a given time (e.g. the wavefront presented in~\cite{He:2021hhl}). It is difficult to apply the spectrum method based on eigenfunctions (e.g. Fourier analysis) in this case.

The perturbation theory of a black hole has already been studied extensively in the literature. Historically, these pioneer works assume that the geometric shape of the hypersurfaces of waves at a given time (wavefronts) is spherical. The angular dependence of waves can then be separated, which leaves a set of master equations along the radius, such as the Regge-Wheeler~\cite{PhysRev.108.1063}, Zerilli~\cite{PhysRevLett.24.737,PhysRevD.2.2141}, Bardeen-Press~\cite{1973JMP....14....7B} equations for the Schwarzschild black hole, and Teukolsky equations~\cite{Teukolsky} for Kerr black hole. Furthermore, by imposing boundary conditions at both the horizon and spatial infinity, the wave equations can be finally solved.

Despite the success of these analyses~\cite{1992mtbh.book.....C}, it is necessary to go beyond the assumption of spherical waves. Even in the weak field limit, as shown in~\cite{He:2021hhl}, when GW signals pass through a compact object, the geometric shape of the outgoing GWs is no longer spherical but is rather complicated~\cite{He:2021hhl}. The conventional analysis based on spherical waves is difficult to describe such situations. New methods are needed. However, due to the complexity of the perturbed equations of a black hole, it is difficult to find analytical expressions for wavefronts with generic shapes. Numerical techniques, therefore, are called for in this case.  

This work aims to extend our work~\cite{He:2021hhl} from simulating the propagation of GWs in a potential well in the weak field limit to the regime of strong gravity. However, a key difference between our work and the scattering theory of black holes~\cite{futterman_handler_matzner_1988,Peters,Suyama:2005mx} is that  our work focuses on a localized wave within a finite spacetime in the time-domain. We treat the propagation of waves as a ``Cauchy" problem for hyperbolic equations, which is fundamentally local. If taking the Fourier transforms, the wave equations in the frequency-domain become elliptic (e.g. Helmholtz equations), which are specific to the ``boundary-value" problem~\cite{nla:cat-vn1414651}. The wave functions, in this case, are
determined by boundary conditions and the shape of the incident wave
packet.  

In this work, we derive the linear perturbation equations of a black hole based on the covariant 3+1 form of Einstein's equations~\cite{gourgoulhon20123}. An advantage of this formalism is that it is less coordinate-dependent, which can avoid the coordinate singularities at $\theta =0\,,\pi$ in the spherical polar coordinates and does not need to impose regularity conditions. This is important for our 3D simulations.

Our evolution scheme is similar to those of solving wave equations in numerical relativity~\cite{PhysRevD.70.104007,PhysRevD.77.084007}. However, a key difference between our work and those presented in~\cite{PhysRevD.70.104007} is that the wave speed in our scheme is no longer a constant but varies in space, which equals the speed of light observed by an asymptotic observer.     

Following our previous work~\cite{He:2021hhl}, the main numerical technique used in this work is called the finite element method (FEM)(see e.g. textbook~\cite{FEMbook} for details). Unlike the conventional numerical method such as the finite difference method (FDM), the FEM is based on the {\it weak formulation} or {\it variational formulation} of partial differential equations (PDEs). The solutions of PDEs can be expanded in terms of {\it ansatz} functions. The domain of interest is then decomposed into finite elements. On each element, a shape function is assigned. If the {\it ansatz} function is the same as the shape function, the scheme is called the Galerkin scheme. The FEM in this case is called the Galerkin FEM. The shape function can be either continuous or discontinuous. The resulting methods are called the continuous Galerkin FEM (cGFEM) or the discontinuous Galerkin FEM (dGFEM)~\cite{osti_4491151}, respectively. Although compared with the cGFEM, the dGFEM is more flexible for the local shape functions and is also more stable for convective problems, it is usually used with an explicit scheme of time discretization and also requires a much larger number of degrees of freedom due to the higher order shape functions used. 

In this work, we adopt the cGFEM method. One particular reason for this is that we implement an implicit scheme of time discretization, which is called the Crank–Nicolson scheme. In flat spacetime, this scheme is symplectic, which inherently preserves the energy of plane waves during their propagation.

Throughout this paper, we adopt the geometric unit $c=G=1$, in which $1\,{\rm Mpc}=1.02938\times 10^{14} {\rm Hz}^{-1}$ and $1 M_{\odot}=4.92535\times 10^{-6} {\rm Hz}^{-1}\,.$ The advantage of this unit system is that time and space have the same unit. 

This paper is organized as follows: In section~\ref{sec::background}, we introduce the 3+1 Einstein's equations for the Schwarzschild spacetime in isotropic coordinates. In section~\ref{sec::perturbed}, we present the perturbed equations up to linear order based on the 3+1 Einstein's equations. In section~\ref{sec::symmetric}, we introduce the symmetric hyperbolic equations for GWs in Schwarzschild spacetime. In section~\ref{sec::FEM}, we introduce the weak formulation of wave equations and the cGFEM method. In section~\ref{sec::numres}, we present our numerical results. In section~\ref{sec::conclusions}, we summarize and conclude this work.

\section{background spacetime\label{sec::background}}
We choose the background as Schwarzschild spacetime. We present the line element in terms of the lapse function $N$ and shift vector $\vec{\beta}$
\begin{equation}
ds^2=g_{\mu\nu}dx^{\mu}dx^{\nu}=-N^2dt^2+\gamma_{ij}(dx^i+\beta^i dt)(dx^j+\beta^j dt)\,\label{lineelement},
\end{equation}
where $\gamma_{ij}$ is the spatial metric. The Greek letters $\mu$ and $\nu$ run from 0 to 3 and the Latin indices $i$ and $j$ run  from 1 to 3. 
In isotropic coordinates, the lapse function and $\gamma_{ij}$ are given by
\begin{align}
N&=\frac{1-\frac{M}{2\rho}}{1+\frac{M}{2\rho}}\,,\label{metric1}\\
\gamma_{ij}&=\left(1+\frac{M}{2\rho}\right)^4\delta_{ij}\,,\label{metric2}
\end{align}
where $\rho=\sqrt{x^ix_i}$ is the radius in isotropic coordinates.
The shift vector vanishes in this case $\vec{\beta}=(0,0,0)$. 

The 3+1 Einstein equations with respect to coordinates $(t,\vec{x})$ are given by~\cite{Baumgarte:2002jm,gourgoulhon20123+1}
\begin{align}
\frac{\partial}{\partial t}\gamma_{ij}&=-2NK_{ij}\,,\label{evolution1}\\
\frac{\partial}{\partial t} K_{ij} &= -D_iD_j N + N (R_{ij}+KK_{ij}-2K_{il} \tensor{K}{^l_j})\label{evolution2}\,,
\end{align}
where $K_{ij}$ is the external curvature tensor, $R_{ij}$ is the Ricci tensor for 3D space and $D_i$ is covariant derivative with respect to the 3D spatial metric $\gamma_{ij}$ . Since $\gamma_{ij}$ is time independent $\frac{\partial}{\partial t}\gamma_{ij}=0$, from Eq.~(\ref{evolution1}) $K_{ij}$ vanishes on all the hyper-surfaces $\Sigma_t$ 
\begin{align}
&K_{ij}=0\,.\label{static0}
\end{align}
The hyper-surface $\sum_t$ in this case is called the maximal slicing $K=0$. 
The lapse functions satisfy the following relations  
\begin{align}
D_iD_jN&=NR_{ij}\neq 0 \,(i\neq j)\,,\label{static1}\\
D_iD^i N&=N R =0\,.\label{static2}
\end{align}

Equations~(\ref{evolution1}) and (\ref{evolution2}) constitute a time evolution system as a Cauchy problem. For the background line elements, inserting Eqs.~(\ref{static0},\ref{static1},\ref{static2}) into Eqs.~(\ref{evolution1}) and (\ref{evolution2}), the background metrics are consistent with Eqs.~(\ref{evolution1}) and (\ref{evolution2}).

In addition to the evolution equations, the 3+1 Einstein's equations are also subject to the Hamiltonian and momentum constraints.
\begin{align}
\mathcal{H}&=\frac{1}{2}\left(R+K^2-K_{ij}K^{ij}\right)=0\,,\label{constraint1}\\
M_i&=D_j\tensor{K}{^j_i}-D_i K = 0\,\label{constraint2}.
\end{align}
For the background spacetime, the above constraints are automatically satisfied  since $R=0$ and $K=0$. 

\section{perturbed spacetime\label{sec::perturbed}}
 
\subsection{perturbed wave equations}
From Eqs.~(\ref{evolution1}) and (\ref{evolution2}), the perturbed 3+1 formalism of the Einstein equations are given by
\begin{align}
\frac{\partial}{\partial t} h_{ij} &= -2\delta N K_{ij}-2N\delta K_{ij}\,,\label{perturbe1}\\
\frac{\partial}{\partial t} \delta K_{ij} &= -\delta (D_iD_jN)+N\delta R_{ij}+N\delta K K_{ij}\nonumber\\
&+N  K \delta K_{ij} -2N\delta K_{il} \tensor{K}{^l_j}-2NK_{il}\delta \tensor{K}{^l_j}\nonumber\\
&+\delta N (R_{ij}+KK_{ij}-2K_{il} \tensor{K}{^l_j})
\,.\label{perturbe2}
\end{align}
where $h_{ij}$ denotes the perturbed metric
\begin{equation}
h_{ij} = \delta \gamma_{ij} \,.
\end{equation}
$\delta R_{ij}$ is the perturbed Ricci tensor,  which can be presented in terms of covariant derivatives $D$ and $\delta \Gamma^{k}_{ij}$ 
\begin{align}
\delta R_{ij}&=D_k\delta \Gamma^k_{ij}-D_j\delta \Gamma^k_{ik}\nonumber\\
&=\frac{1}{2}(D^lD_i h_{lj}+D^lD_jh_{il}-D^lD_lh_{ij})\nonumber\\
&-\frac{1}{2}D_jD_i(\gamma^{kl}h_{lk})\,.\label{deltaRicci}
\end{align}
Although the Christoffel symbol $\Gamma^{k}_{ij}$ itself is not a tensor, its perturbation 
 $\delta \Gamma^{k}_{ij}$ is a tensor, which can be written in a covariant format~\cite{PhysRev.146.938}
\begin{align}
\delta \Gamma^{k}_{ij} =\frac{1}{2}\gamma^{kl}(D_i h_{lj}+D_j h_{il}-D_l h_{ij})\,.
\end{align}
The perturbation of $2N\delta(D_iD_jN)$ is given by
\begin{align}
2N\delta(D_iD_jN)&=2N\left(D_iD_j\delta N-\delta \tensor{\Gamma}{^k_i_j}\partial_k N\right) \nonumber\\
&=2N\left(\frac{\partial^2 \delta N}{\partial x^i\partial x^j}
-\tensor{\Gamma}{^k_i_j}\partial_k\delta N\right)\nonumber\\
&-N\partial_kN\gamma^{kl}(D_ih_{lj}+D_jh_{il}-D_lh_{ij})\,.\label{deltaN}
\end{align}
where
\begin{align}
D_iD_j\delta N=\frac{\partial^2 \delta N}{\partial x^i\partial x^j}-\tensor{\Gamma}{^k_i_j}\partial_k\delta N\,.
\end{align}

Next, taking the time derivative of Eq.~(\ref{perturbe1}),
and using Eq.~(\ref{perturbe2}) to eliminate $\delta K_{ij}$,  we obtain a second order equation for $h_{ij}$
\begin{align}
\frac{\partial^2}{\partial t^2}h_{ij}&=-2\frac{\partial \delta N}{\partial t} K_{ij}-2\delta N\frac{\partial K_{ij}} {\partial t} \nonumber\\
&+2N\delta(D_iD_jN)-2N^2\delta R_{ij}\nonumber\\
&-2N^2\delta K K_{ij}-2N^2K\delta K_{ij} + 4N^2\delta K_{il}\tensor{K}{^l_j} \nonumber\\
&+4N^2 K_{il}\delta \tensor{K}{^l_j}-2N\delta N (R_{ij}+KK_{ij}-2K_{il} \tensor{K}{^l_j})
\,. \label{waveequation}	
\end{align}
Noting that the background quantities vanish $K_{ij}=0$ $K=0$ and $\tensor{K}{_i^j}=0$, the above equation reduces to
\begin{align}
\frac{\partial^2}{\partial t^2}h_{ij}&= 2N\delta(D_iD_jN)-2N^2\delta R_{ij}-2N\delta N R_{ij}\,.\label{waveequation2} 
\end{align}
Inserting Eq.~(\ref{deltaRicci}) and Eq.~(\ref{deltaN}) into Eq.~(\ref{waveequation}), we obtain
\begin{align}
\frac{\partial^2}{\partial t^2}h_{ij}&=2ND_iD_j\delta N-2N\delta N R_{ij}\nonumber \\
+&N^2(D_iD_jh+D^lD_lh_{ij}-D^lD_ih_{lj}-D^lD_jh_{il})\nonumber\\
-&N\partial_kN\gamma^{kl}(D_i h_{lj}+D_j h_{il}-D_l h_{ij})\,.
\end{align}
Using the identity
\begin{align}
D^lD_ih_{lj}=D_iD^l h_{lj}+\tensor{R}{^m_j_i^n}h_{mn}+\tensor{R}{_i^l}h_{lj}\,,
\end{align}
we finally arrive at
\begin{align}
\frac{\partial^2}{\partial t^2}h_{ij}&=2ND_iD_j\delta N-2N\delta N R_{ij}\nonumber \\
&+N^2D^lD_lh_{ij}+N^2D_iD_jh-N^2\left(D_i \Gamma_j+D_j\Gamma_i\right)\nonumber \\
&-N^2(2\tensor{R}{^m_i_j^n}h_{mn}+\tensor{R}{_i^l}h_{lj}+\tensor{R}{_j^l}h_{li})\nonumber\\
&-N\partial_kN\gamma^{kl}(D_i h_{lj}+D_j h_{il}-D_l h_{ij})\,, \label{wavetensor}
\end{align}
where
\begin{align}
\Gamma_j=D^{l}h_{lj}\,.
\end{align}

Equation~(\ref{wavetensor}) gives the most general linear perturbed equation for gravitational waves in Schwarzschild spacetime. If $\gamma_{ij}$ is in flat case, Eq.~(\ref{wavetensor}) is consistent with Eq.(85) presented in Ref.~\cite{PhysRevD.70.104007}. 

\subsection{general covariance}
Unlike the non-linear perturbations, in linear perturbation theory general covariance plays a vital role. To highlight this point, we consider an arbitrary infinitesimal coordinate transformation $\eta^i$
\begin{equation}
\tilde{\rho}^{i}=\rho^{i}+\eta^i\,.
\end{equation} 
For an scalar field $S$ and a tensor field $T_{ij}$, the perturbed quantities transform under $\eta^i$ as 
\begin{align}
\delta\tilde{S} &\rightarrow \delta S-\mathcal{L}_{\vec{\eta}}S\,,\\	
\delta\tilde{T}_{ij} &\rightarrow \delta T_{ij}-\mathcal{L}_{\vec{\eta}}T_{ij}\,,
\end{align}
where $\mathcal{L}_{\vec{\eta}}$ denotes the Lie derivative. For a scalar field, the Lie derivative gives
\begin{equation}
\mathcal{L}_{\vec{\eta}}S=  \eta^kD_k S\,,
\end{equation}
and for a tensor field, the Lie derivative reads
\begin{equation}
\mathcal{L}_{\vec{\eta}}T_{ij}=T_{ik}D_j\eta^k+T_{kj}D_i\eta^k + \eta^kD_k T_{ij}\,.
\end{equation}
The perturbed quantities $\delta N,\,$ $h_{ij}$ and $\delta R_{ij}$ then transform as
\begin{equation}
\left\{
\begin{aligned}
\delta \tilde{N}&\rightarrow \delta N - \eta^kD_k N\\
\tilde{h}_{ij}&\rightarrow h_{ij} - \gamma_{ik}D_j\eta^k-\gamma_{kj}D_i\eta^k - \eta^kD_k\gamma_{ij}\\
\delta \tilde{R}_{ij}&\rightarrow \delta R_{ij} - R_{ik}D_j\eta^k-R_{kj}D_i\eta^k - \eta^kD_k R_{ij} \label{infinitrans}
\end{aligned} \right.\,.
\end{equation}

Since the gauge transformation simply changes coordinates, it does not induce real  physics. If an equation describes real physics, it should have the same format in different coordinates. This principle is known as general covariance. It turns out that Eq.~(\ref{waveequation2}) does have such a property. When inserting the above gauge transformation,  Eq.~(\ref{waveequation2}) keeps the same format in the new coordinate system. The detailed proof is provided in appendix~\ref{covariantproof}. A key point in the proof is that $\delta N$ does not vanishes, which plays an essential role for Eq.~(\ref{waveequation2}) to be covariant.  

If $\sum_t$ is compact or $\sum_t$ is open but $\eta^i$ can vanish rapidly at infinity, the tensor field $h_{ij}$ can be decomposed into scalar, vector and tensor components according to their properties of gauge transformation \cite{1984Psasaki}. However, although in most cases it is possible to do such decomposition for $h_{ij}$, there is no guarantee that different components can evolve independently. This is because the evolution equations of different components may couple to one another. To illustrate this point, we decompose $\eta^k$ as
\begin{align}
&\eta^k = D^k\eta + \eta^k_* \,,\nonumber\\
&D_k\eta^k_*=0\,.
\end{align}
$\eta$ and $\eta^k_*$ then represent the scalar and vector infinitesimal transformations, respectively.
Note that these two transformations are independent to each other. $\eta^k_*$ alone, therefore, does not change the scalar components of $h_{ij}$. However, this may not be true for the second order derivatives of $h_{ij}$
\begin{align}
D^iD^j\tilde{h}_{ij}\rightarrow D^iD^j h_{ij} - 2 R_{ij}D^i\eta^j_*\,.
\end{align}  
Although $D^iD^j h_{ij}$ itself is a scalar field, whether it will be changed under the vector coordinate transformation $\eta^k_*$ is dependent on the curvature of $\sum_t$. 
If $R_{ij}$ is to be maximally symmetric, such as in the case of Robertson-Walker metric
\begin{align}
R_{ij}=2K\gamma_{ij}\,,
\end{align} 
$R_{ij}D^i\eta^j_*$ then vanishes. The vector transformation $\eta^k_*$ in this case does not induce any change in the scalar quantity $D^iD^j h_{ij}$. 
And the equations for scalar and vector components can evolve independently. In cosmological perturbation theory, the explicit proof of the independence of different components can be found in the Appendix B of Ref.~\cite{1984Psasaki} where the constant curvature of the background space plays an essential role in the proof. 

However, in our case, $R_{ij}D^i\eta^j_*$ does not vanish since $R_{ij}\neq 0 \, (i\neq j)$ and $R_{ij}$ can not be presented as a constant in combination with $\gamma_{ij}$. Different components in Eq.~(\ref{wavetensor}), thus, couple to one another. In this case, the most general perturbed equation should be used directly.

\subsection{perturbed conservation laws\label{pconser}}
In addition to the wave equation, the perturbed constraint Eq.~(\ref{constraint1}) gives
\begin{align}
\delta R + 2K\delta K - \delta K_{ij}K^{ij}-K_{ij} \delta K^{ij} = 0\,.
\end{align}
Since the background external curvature vanishes, we obtain
\begin{align}
\delta R = 0\,.
\end{align}

From Eq.~(\ref{deltaRicci}), we obtain
\begin{align}
\gamma^{ij} \delta R_{ij} = D^l \Gamma_l-D^lD_l h\,.
\end{align}
Further noting that
\begin{align}
\delta R = \delta \gamma^{ij} R_{ij} + \gamma^{ij} \delta R_{ij}
\end{align} 
and
\begin{align}
\delta \gamma^{ij} = -\gamma^{im}\gamma^{jn} h_{mn}\,,
\end{align}
we obtain
\begin{align}
D^l \Gamma_l-D^lD_l h=\gamma^{im}\gamma^{jn} h_{mn}R_{ij}\,.\label{Dgamma}
\end{align}
The perturbation of Eq.~(\ref{static2}) gives
\begin{align}
\delta(D_iD^i N)=\gamma^{ij}\delta(D_iD_jN)-\gamma^{im}\gamma^{jn}h_{mn}D_iD_jN=0
\end{align}
Contracting Eq.~(\ref{waveequation2}) with $\gamma^{ij}$, we obtain
\begin{align}
\frac{\partial^2h}{\partial t^2}&=2N\gamma^{ij}\delta(D_iD_j N)-2N^2\gamma^{ij}\delta R_{ij}\nonumber\\
&=2N\gamma^{im}\gamma^{jn}h_{mn}(D_iD_j N - N R_{ij})\nonumber \\
&=0\label{wavetrace}
\end{align}
Therefore, we can take the trace of $h_{ij}$ as zero $h=0$.
Equation~(\ref{Dgamma}), then, gives
\begin{align}
D^l\Gamma_l=\gamma^{im}\gamma^{jn}h_{mn}R_{ij}=\gamma^{im}\gamma^{jn}h_{mn}\frac{D_iD_jN}{N}\,.\label{Gamma_new}
\end{align}

Next, contracting Eq.~(\ref{perturbe1}) with $\gamma^{ij}$ and noting that $h=0$, we obtain
\begin{align}
\delta K = 0\,.
\end{align}
The perturbation of momentum constraint Eq.~(\ref{constraint2}) reduces to
\begin{align}
D^j \delta K_{ji} = 0\,.\label{mconstraint}
\end{align}
Taking the spatial derivative of Eq.~(\ref{waveequation2}), we obtain
\begin{align}
\frac{\partial}{\partial t} D^i h_{ij} = -2D^iN\delta K_{ij}-2ND^i\delta K_{ij}\,.
\end{align}
Using the definition
\begin{align}
	\Gamma_j=D^{l}h_{lj}
\end{align}
and Eq.~(\ref{mconstraint}), we find
\begin{align}
\frac{\partial}{\partial t} \Gamma_j = D^i\ln N\frac{\partial}{\partial t} h_{ij}\,.
\end{align}
The above equation has a general solution, which takes the form
\begin{align}
\Gamma_l = D^m \ln N h_{ml} +A_l \,,\label{Gammahml}
\end{align}
where $A_l$ is a free vector field that does not depend on time $\frac{\partial {A_l}}{\partial t}=0$.

To fix the choice of the vector field $A_l$, we can take the spatial derivative of the Eq.~(\ref{Gammahml})
\begin{align}
D^l\Gamma_l &= (D^l D^m \ln N) h_{ml} + D^m \ln N D^l h_{ml} +D^l A_l \nonumber\\
&=(D^n D^m \ln N + D^m \ln N D^n \ln N) h_{mn} \nonumber\\
&+ D^m \ln N A_m + D^m A_m \nonumber\\
&= \frac{D^mD^n N}{N} h_{mn}\,. \label{Gammaequality}
\end{align}
In the last equality, we have used Eq.~(\ref{Gamma_new}).
Further using the identity
\begin{align}
D^n D^m \ln N + D^m \ln N D^n \ln N = \frac{D^mD^n N}{N}\,,
\end{align}
from Eq.~(\ref{Gammaequality}) we obtain
\begin{align}
D^m \ln N A_m + D^m A_m = 0\,.
\end{align}
The above equation places a constraint on the choice of $A_m$. In this work, we take $A_m = 0$, which is an obvious solution to the above equation. Equation~(\ref{Gammahml}) reduces to 
\begin{align}
\Gamma_l = D^m \ln N h_{ml} \,.\label{GammahmlNOa}
\end{align}
The above relation plays a vital role in the numerical process of solving Eq.~(\ref{wavetensor}). We will discuss this point in the next few sections.
\subsection{Choosing coordinates}
As already noted, an infinitesimal coordinate transformation from the background spacetime can lead to non-zero perturbations. Although these perturbations still satisfy the covariant perturbed equation Eq.~(\ref{waveequation2}), it is obvious that they do not represent any true physics.

One way to get around this problem is to evolve the perturbed equations in a particular gauge and then extract the gauge-invariant GWs from the symmetric trace-free tensor $h_{ij}$ later on ( see ~\cite{gourgoulhon20123+1} for reviews). Choosing a gauge is usually done by picking up particular forms for the lapse $N$ and shift $\beta_i$. In our work, since $\beta_i=0$, we only need to choose a particular form for $N$. In this work, we choose a gauge in which $\delta N =0$. This slicing is known as the maximal slicing since the mean external curvature $K=\delta K=0$ vanishes on this hypersurfaces $\Sigma_t$.

\section{symmetric hyperbolic equations\label{sec::symmetric}}
For numerical reasons, it is more convenient to write the covariant derivatives $D^lD_l$ in Eq.~(\ref{wavetensor}) in terms of ordinary partial derivatives 
\begin{align}
N^2D^lD_lh_{ij}=c^2\nabla^2 h_{ij}-\frac{32(2\rho-M)^2\rho^3M}{(2\rho+M)^7}\frac{\rho^l}{\rho}\partial_{l} h_{ij}\,,
\end{align}
where $\nabla^2 = \partial_i\partial^i$ and the coefficient $c^2$ is defined by
\begin{align}
c^2 &= \frac{16\rho^4 (2\rho-M)^2}{(2\rho+M)^6}\,.\label{definationspeed}
\end{align}
$c$ has a clear physical meaning, which is the speed of GWs in Schwarzschild spacetime. Note that $c$ is not a constant but varies in space.

In fact, it has long been known that the major challenge of solving the wave equation Eq.~(\ref{wavetensor}) lies in the terms associated with $\partial_i \Gamma_j$, which involve the mixed second order spatial derivatives of $h_{ij}$~\cite{Baumgarte:2010ndz}. The wave equation Eq.~(\ref{wavetensor}) in this case is no longer {\it symmetric hyperbolic} and {\it well-posed}, as
the mixed terms can generate modes that grow exponentially in the solutions~\cite{Baumgarte:2010ndz}.  One way to overcome this problem is to present $\Gamma_j$ in terms of $h_{ij}$ rather than its spatial derivatives using the momentum constraint Eq.~(\ref{GammahmlNOa})
\begin{align}
\Gamma_j = \gamma^{mn}D_n \ln N h_{mj}=\frac{f_{\Gamma}}{N^2}\frac{\rho^m}{\rho} h_{mj}\,.
\end{align}
We then obtain
\begin{align}
&N^2\left(\partial_i \Gamma_j+\partial_j\Gamma_i-2\tensor{\Gamma}{^k_i_j}\Gamma_k\right)\nonumber \\
=&f_{\kappa}\frac{\rho^m\rho^k}{\rho^2}(\delta_{ki}h_{mj}+\delta_{kj}h_{mi})\nonumber \\
+&f_{\Gamma}\frac{\rho^l}{\rho}\left(\partial_i h_{lj}+\partial_j h_{il}-2 \tensor{\Gamma}{^m_i_j} h_{lm}\right)\,,\label{N2Gamma}
\end{align}
where 
\begin{align}
f_{\kappa}&=-\frac{64M\rho^3(3M^2-8M\rho+12\rho^2)}{(2\rho+M)^8}\nonumber\\
f_{\Gamma}&= \frac{64(2\rho-M) M\rho^4}{(2\rho+M)^7}\,.
\end{align}
Equation~(\ref{N2Gamma}), thus, does not involve the mixed second order spatial derivatives of $h_{ij}$. The wave equation, in this case, is {\it symmetric hyperbolic} and is also {\it strong hyperbolic} (see Chapter 11.1 in~\cite{Baumgarte:2010ndz}). The wave equation is {\it well-posed} and turns out to be stable in the numerical process.

Next, we rewrite the Ricci tensor in the second term on the RHS in Eq.~(\ref{wavetensor}) as
\begin{align}
N^2\gamma^{lk}\tensor{R}{_i_k}h_{lj}=f_R\delta^{lk}\tilde{R}_{ik}h_{lj}\,,
\end{align}
where $R_{ij}$ and $\tilde{R}_{ij}$ are given by
\begin{align}
R_{ij}&=\frac{4M}{\rho^3(2\rho+M)^2}(\delta_{ij}\rho^2-3\rho_i\rho_j)\nonumber\\
&=\frac{4M}{\rho^3(2\rho+M)^2}\tilde{R}_{ij}\,,\\
\tilde{R}_{ij}&=\delta_{ij}\rho^2-3\rho_i\rho_j\,.
\end{align}
The coefficient $f_{R}$ is defined by
\begin{align}
f_R = \frac{64(2\rho-M)^2 M\rho}{(2\rho+M)^8}\,.
\end{align}
The terms associated with the Riemann tensor can be written as
\begin{align}
N^2\tensor{R}{^m_i_j^n}h_{mn}=N^2\gamma^{mp}\gamma^{nq}\tensor{R}{_p_i_j_q}h_{mn}\nonumber\\
=f_R\delta^{mp}\delta^{nq}\tensor{\tilde{R}}{_p_i_j_q}h_{mn}\,,
\end{align}
where 
\begin{align}
R_{pijq}&=\frac{M(2\rho+M)^2}{4\rho^7}\tilde R_{pijq}\,,\\
\tilde R_{ijij}&=2\rho^2-3\rho_i^2-3\rho_j^2\,,\quad(i\neq j)\\
\tilde {R}_{ijik}&=-3\rho_j\rho_k\,,\quad(i\neq j \neq k)\,.
\end{align}

The last term on the RHS in Eq.~(\ref{wavetensor}) can be simplified as
\begin{align}
&N\partial_kN\gamma^{kl}(D_i h_{lj}+D_j h_{il}-D_l h_{ij})\nonumber\\
=&f_{\Gamma}\frac{\rho^l}{\rho}\left[\partial_i h_{lj}+\partial_j h_{il}-\partial_l h_{ij}-2 \tensor{\Gamma}{^m_i_j} h_{lm}\right]\,.
\end{align}

Finally, combining all the above expressions, we obtain
\begin{align}
\frac{\partial^2}{\partial t^2}h_{ij}&=c^2\nabla^2 h_{ij}-f_{\kappa}\frac{\rho^m\rho^k}{\rho^2}(\delta_{ki}h_{mj}+\delta_{kj}h_{mi})\nonumber\\
&+f_{\rho}\frac{\rho^k}{\rho}\partial_{k} h_{ij}-2f_R\delta^{mp}\delta^{nq}\tensor{\tilde{R}}{_p_i_j_q}h_{mn}\nonumber\\
&-f_R(\delta^{lk}\tilde{R}_{ik}h_{lj}+\delta^{lk}\tilde{R}_{jk}h_{li})\nonumber\\
&-2f_{\Gamma}\frac{\rho^l}{\rho}\left(\partial_i h_{lj}+\partial_j h_{il}-2 \tensor{\Gamma}{^m_i_j} h_{lm}\right)\,,\label{perturbedwaveequation}
\end{align}
where 
\begin{align}
f_{\rho}&=\frac{32(2\rho-M)\rho^3M^2}{(2\rho+M)^7}\nonumber\,. 
\end{align}
The above equation is the core equation we aim to solve in this work. Before going further, we shall address several key aspects of Eq.~(\ref{perturbedwaveequation}).

\subsection{isotropic wave speed}
The principle part of the wave equation Eq.~(\ref{perturbedwaveequation}) is
\begin{align}
\frac{\partial^2}{\partial t^2}h_{ij}&=c^2\nabla^2 h_{ij}+{\rm dispersion\,terms} \,,
\end{align}
where $c$ is the speed of GWs measured by a static observer $(\frac{\partial}{\partial t})^a$ at spatial infinity. Its value varies in space. 
The rest terms in Eq.~(\ref{perturbedwaveequation}) serve as dispersion terms. If Eq.~(\ref{perturbedwaveequation}) admits a plane wave solution, these terms change the dispersion relation between the speed of phase and the frequency of the wave.

\begin{figure}
{\includegraphics[width=\linewidth]{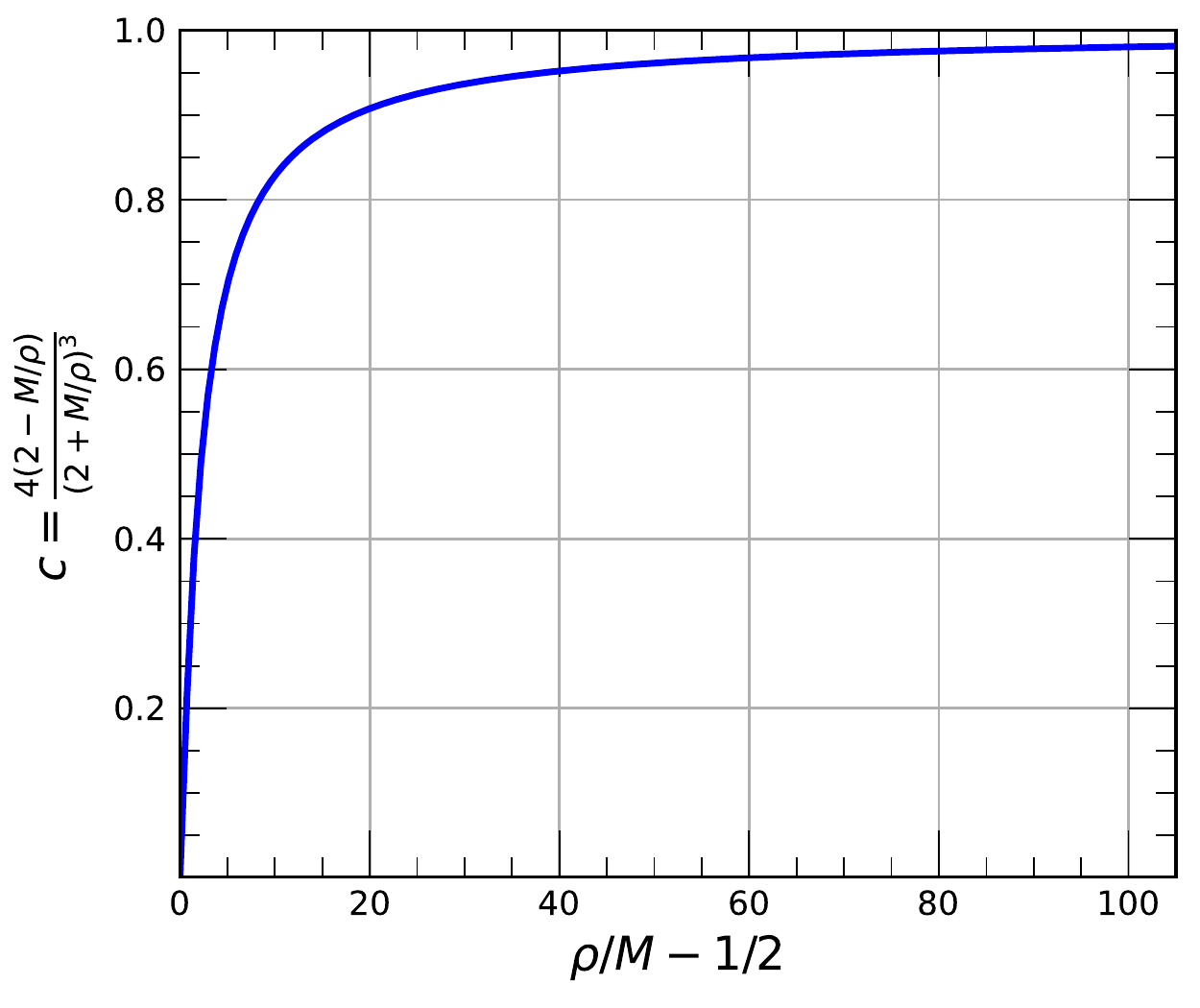}}
\caption{The speed of wave $c$ as a function of $\rho/M$. We choose the horizontal axis as $\rho/M-1/2$ for convenience. When observers are far away from the black hole or the mass of the black hole becomes zero, $c$ goes back to the speed of light in vacuum. Conversely, at the horizon $\rho/M\rightarrow 1/2$, the speed of wave becomes zero.
\label{wavespeed}}
\end{figure}

Figure~\ref{wavespeed} shows $c$ as a function of $\rho/M$. If it is far away from the black hole
\begin{equation}
\lim_{\rho/M\rightarrow \infty}c = 1\,,
\end{equation}
$c$ goes back to the speed of light in vacuum $c\sim 1$. 
At the horizon $c$ becomes zero
\begin{equation}
\lim_{\rho\rightarrow M/2}c = 0\,.
\end{equation}

Moreover, it is important to note that $c$ equals the speed of light in Schwarzschild spacetime. To see this point, we consider a null curve $ds^2=0$ (not necessarily null geodesics). From the line element Eq.~(\ref{lineelement}), we obtain
\begin{align}
0=-N^2dt^2+\left(1+\frac{M}{2\rho}\right)^4dl^2\,,
\end{align}
where $dl^2=dx^2+dy^2+dz^2$.
The above equation gives
\begin{align}
\frac{dl^2}{dt^2}=\frac{N^2}{\left(1+\frac{M}{2\rho}\right)^4}=\frac{16\rho^4 (2\rho-M)^2}{(2\rho+M)^6}=c^2\,.\label{wavespeed_light_rays}
\end{align}
This demonstrates that GWs in Schwarzschild spacetime travel at the same speed as those of light rays. Moreover, it is also worth noting that $c$ is isotropic in our coordinate system, namely, the speed of wave does not depend on the direction of propagation.  

\subsection{Flat spacetime limit}
Next, we demonstrate  the consistency of our formalism in the limit of flat spacetime, namely,  $M\rightarrow 0$ and $N\rightarrow 1$. Here we also assume that $\delta N\rightarrow 0$.

In this case, Eqs.~(\ref{perturbe1},\ref{perturbe2},\ref{waveequation}) reduce to 
\begin{align}
\frac{\partial}{\partial t} h_{ij} &= -2\delta K_{ij}\,,\\
\frac{\partial}{\partial t} \delta K_{ij} &= \delta R_{ij},\\
\frac{\partial^2}{\partial t^2}h_{ij}&=-2\delta R_{ij}\,. \label{flatwaveequation}	
\end{align}
From Eq.~(\ref{deltaRicci}), the perturbed Ricci tensor becomes
\begin{align}
\delta R_{ij}=-\frac{1}{2}\nabla^2 h_{ij}\,.\label{flatDeltaR}
\end{align}
Equation~(\ref{waveequation}) then becomes
\begin{align}
\frac{\partial^2}{\partial t^2}h_{ij}=\nabla^2 h_{ij}\,.\label{planewave}
\end{align}
The above equation can also be obtained from Eq.~(\ref{perturbedwaveequation}) by noting that $f_{\rho}\rightarrow 0\,,f_{R}\rightarrow 0\,,f_{\Gamma}\rightarrow 0 \,, c^2\rightarrow 1$ when $M\rightarrow 0$.

Equation~(\ref{planewave}) is consistent with the well-known wave equation in Minkowski spacetime. Note that, in the flat spacetime limit, the momentum constraint gives $\Gamma_i \rightarrow 0$. 

Equation~(\ref{planewave}) has an analytical solution
\begin{align}
h_{ij}(t,\vec{\rho})=H_{ij} \cos(\omega t -k^i\rho_i)\,,\label{planewaves}
\end{align}
where $\vec{k}$ is the wave vector, $\vec{\rho}$ is the position vector and $\omega$ is the angular frequency. $H_{ij}$ is a constant tensor field. Taking the derivative $\nabla^i$ of $h_{ij}$, the condition $\Gamma_j=\nabla^ih_{ij}=0$ gives
\begin{align}
\nabla^ih_{ij}=k^i H_{ij} =0\,.
\end{align}
The above equation, thus, indicates that $H_{ij}$ is perpendicular to the wave vector $\vec{k}$, which means that the oscillation of GWs is  transverse relative to the direction of its propagation. Therefore, in flat spacetime, the condition $\Gamma_j=0$ implies transverse waves.

Moreover, since $h_{ij}$ is trace-less, from Eq.~(\ref{flatDeltaR}) we obtain
\begin{align}
\delta R = \delta \gamma^{ij}R_{ij}+\delta\tensor{R}{^i_i}=\delta\tensor{R}{^i_i}=-\frac{1}{2}\nabla^2 {\tensor{h}{^i_i}}=0\,.
\end{align} 
Given the transverse condition, we also have
\begin{align}
\nabla^i\delta K_{ij}=-\frac{1}{2}\frac{\partial}{\partial t} \nabla^i h_{ij} =0\,.
\end{align}
As such, Eq.~(\ref{planewaves}) satisfies both the energy and momentum constraints. 

It is worth noting that, in general, $\Gamma_j$ does not vanish in curved spacetime, unless the GW tensor is transverse relative to the radius.
This is because from the momentum constraint 
\begin{align}
\Gamma_l = D^m \ln N h_{ml} \propto \rho^m h_{ml},
\end{align}
a vanishing $\Gamma_l$ leads to  a transverse GW tensor relative to the radius $\rho^m h_{ml}=0$. This usually happens for spherical waves. However, in our case GWs are neither plane waves nor spherical waves. $\Gamma_l$, therefore, does not vanish.

\subsection{wave equations at the horizon}
At the horizon $\rho \rightarrow M/2$, the only non-vanishing coefficient in Eq.~(\ref{perturbedwaveequation}) is
\begin{align}
f_{\kappa}\rightarrow-\frac{1}{16M^2}\,.
\end{align}
Equation~(\ref{perturbedwaveequation}) then reduces to
\begin{align}
\frac{\partial^2}{\partial t^2}h_{ij}&=\frac{\rho^m\rho^k}{4M^4}(\delta_{ki}h_{mj}+\delta_{kj}h_{mi})\,,
\end{align}
where $\rho^k\rho_k=M^2/4$. It is worth noting that there is no divergent terms in our formulation. 

If the incident waves are of spherical symmetry and transversely relative to the radius $\rho^mh_{mi}=0$,
the above equation simply gives $h_{ij}(t)|_{\rho=M/2}=0$, given that the horizon is initially at rest. This is reasonable since the transverse GWs only induce oscillations that lie on the surface of the horizon, which do not actually change the horizon.
  
Only when GWs have non-transverse components, the above equation has a non-trivial solution. For instance, if the GW tensor has a non-zero component $h_{xx}$ and hits the horizon at $(-M/2,0,0)$, the above equation reduces to
\begin{align}
\frac{\partial^2}{\partial t^2}h_{xx}&=\frac{1}{8M^2}h_{xx}\,,
\end{align}   
which admits a decaying mode in the solution $h_{xx} =C e^{-\frac{\sqrt{2}}{4M}t}$. This indicates that GWs are not frozen at the horizon but
will die out with the asymptotic time $t$. However, it should be noted that since the wave speed at the horizon is zero $c \rightarrow 0$, no information at the horizon can propagate to a distant observer. This is consistent with the usual treatment by pushing the event horizon of the black hole to $-\infty$ using the tortoise coordinate, in which information at the horizon takes infinite asymptotic time to get out of the horizon. Our treatment, indeed, achieves a similar result but with a more natural geometry for the black hole in 3D space.

\section{Finite Element method\label{sec::FEM}}
To numerically solve Eq.~(\ref{perturbedwaveequation}), we use the finite element method. Unlike the conventional methods such as the finite difference method, the FEM is based on the {\it weak formulation} (or {\it variational formulation}) of the PDEs. Therefore we will first introduce the {\it weak formulation} of the wave equations, which can be obtained by multiplying Eq.~(\ref{perturbedwaveequation}) with a test function $\Psi$ and then integrating over a domain $\Omega$. For convenience, we adopt the following notion for short
\begin{equation}
\langle f,g \rangle=\int_{\Omega}f(x)^*g(x)\,\mathrm{d}x\,.
\end{equation}
Equation~(\ref{perturbedwaveequation}) can be presented as
\begin{align}
&\langle\Psi,\frac{\partial^2}{\partial t^2}h_{ij}e^i\otimes e^j\rangle\nonumber\\
=&\langle\Psi, (c^2\nabla^2 h_{ij}+f_{\rho}\partial_{\rho} h_{ij})e^i\otimes e^j\rangle \nonumber\\
-&\langle\Psi, f_{\kappa}\frac{\rho^m\rho^k}{\rho^2}(\delta_{ki}h_{mj}+\delta_{kj}h_{mi})e^i\otimes e^j\rangle \nonumber\\
-&\langle\Psi, (f_R\delta^{mp}\tilde{R}_{im}h_{pj}+f_R\delta^{mp}\tilde{R}_{jp}h_{mi})e^i\otimes e^j\rangle  \nonumber\\
-&2\langle\Psi,f_R\delta^{mp}\delta^{nq}\tensor{\tilde{R}}{_p_i_j_q}h_{mn}e^i\otimes e^j\rangle \nonumber\\
-&2\langle\Psi,f_{\Gamma}\frac{\rho^p}{\rho}\left(\partial_i h_{pj}+\partial_j h_{ip}-2 \tensor{\Gamma}{^m_i_j} h_{pm}\right)e^i\otimes e^j\rangle \,, \label{weakwaveequation} 
\end{align}
where $e^i\otimes e^j$ is the tensor basis. As pointed out previously, $h_{ij}$ is trace-less. Due to symmetry, $h_{ij}$ only has $5$ independent components. We denote the basis for these $5$ independent components as $\epsilon^{\alpha}$, which is related to $e^i\otimes e^j$ by
\begin{align}
e^i\otimes e^j=\tensor{C}{^i^j_{\alpha}}\epsilon^{\alpha}\,.\label{tensorbasis}
\end{align}
The non-vanishing components of $\tensor{C}{^i^j_{\alpha}}$ are
\begin{align}
&\tensor{C}{^2^2_{0}}=\tensor{C}{^1^1_{2}} = 1\nonumber\,,\\
&\tensor{C}{^2^3_{1}}=\tensor{C}{^3^2_{1}} = \tensor{C}{^2^1_{3}}= \tensor{C}{^1^2_{3}}= \tensor{C}{^3^1_{4}}= \tensor{C}{^1^3_{4}}=1/2\,.
\end{align}
Given such a basis, the tensor field $h_{ij}e^i\otimes e^j$ can be presented in terms of $\epsilon^{\sigma}$
\begin{align}
h_{ij}e^i\otimes e^j=H_{\sigma}\epsilon^{\sigma}\,.
\end{align}
The components of $H_{\sigma}$ are explicitly given by
\begin{equation}
\left\{
\begin{aligned}
&H_0=h_{22}\nonumber\\
&H_1=h_{\times}=h_{23}=h_{32}\nonumber\\
&H_2=h_{11}\nonumber\\
&H_3=h_{12}=h_{21}\nonumber\\
&H_4=h_{13}=h_{31}\nonumber\\
&h_{33}=-H_2-H_0
\end{aligned} \right.\,.
\end{equation}
$H_{\sigma}$ is then related to $h_{ij}$ by 
\begin{align}
h_{ij}=\tensor{C}{_i_j^{\sigma}}H_{\sigma}\,.
\end{align}
The non-vanishing components of $\tensor{C}{_i_j^{\sigma}}$ are
\begin{align}
\tensor{C}{_2_2^{0}}&=\tensor{C}{_2_3^{1}}=\tensor{C}{_3_2^{1}}=\tensor{C}{_1_1^{2}} \nonumber\\
&=\tensor{C}{_2_1^{3}}=\tensor{C}{_1_2^{3}}=\tensor{C}{_2_1^{4}}=\tensor{C}{_1_3^{4}}=1\,,\nonumber \\
\tensor{C}{_3_3^{0}}&=\tensor{C}{_3_3^{2}}=-1\,.
\end{align}
Note that $\tensor{C}{_i_j^{\sigma}}$ and $\tensor{C}{^i^j_{\sigma}}$ are symmetric with respect to $i\,,j$.
They are related to each other by 
\begin{align}
\tensor{C}{_i_j^{\sigma}}\tensor{C}{^i^j_{\alpha}}=\tensor {\delta}{^{\sigma}_{\alpha}}\,.
\end{align}

The {\it test} function $\Psi$ for a vector field can be constructed in a form
\begin{align}
\Psi = \phi \otimes \epsilon^{\alpha}\,,
\end{align}
where $\phi$ is chosen in such a way that it vanishes on the subset of boundaries with Dirichlet boundary conditions ${\partial \Omega_D}$
\begin{equation}
\mathbb{V}:=\{\phi:\phi\in \mathbb{H}^{1}(\Omega),\phi|_{\partial \Omega_D}=0\}\,.\label{Vspace}
\end{equation}
$\mathbb{H}^{1}(\Omega)=\mathbb{W}^{1,2}(\Omega)$ is called the first order {\it Sobolev space} meaning that $\phi$ and its first order weak derivatives $\partial_x \phi$ are square integrable
\begin{equation}
\left\|\phi\right\|_{\mathbb{H}^1(\Omega)}=\left[\int_{\Omega}\sum_{|\alpha|\leq 1} |\partial^{\alpha}_x\phi(x)|^2dx\right]^{\frac{1}{2}}<\infty\,.\nonumber
\end{equation}

If $h_{ij}$ in Eq.~(\ref{weakwaveequation}) holds for any test function $\Psi$, $h_{ij}$ is called the {\it weak solution} and Eq.~(\ref{weakwaveequation}) is called the {\it weak formulation}.
\subsection{spatial discretization}
In the Finite Elements Methods (FEM), the domain $\Omega$ is decomposed into subdomains $\Omega_i$, which consist of rectangles or triangles. This is called {\it decomposition} or {\it triangulation}. The vertices of rectangles and triangles in the domain $\Omega$ are called mesh points or nodes. Let $\Omega_h$ denote the set of all nodes of the decomposition. On each node, we construct a scalar test function $\phi_i\in \mathbb{V} \,,i=1,..,N$, where $\mathbb{V}$ is the space defined in Eq.~(\ref{Vspace}) and $N$ is the total number of nodes in the domain. The scalar test function $\phi_i$ is required to have the property
\begin{equation}
\phi_i(p^k)=\delta_{ik}, \quad i,k=1,..,N, \quad p^k\in \Omega_h \,. \nonumber
\end{equation}

As such, $\phi_i$ has non-zero values only on the node with $k=i$ and its adjacent subdomains, which is called the influential zones. However, it vanishes on other parts of the domain $\Omega$. The test function $\phi_i$ constructed this way is called the {\it scalar shape function}. Clearly, $\phi_i \in \mathbb{V}$ on different nodes are linearly independent. We denote the space spanned by $\phi_i$ as
$\mathbb{V}_h:=\mathrm{span}\{\phi_i\}_{i=1}^N$, which is a subspace of $\mathbb{V}$. The vector shape functions can be constructed from the scalar shape functions with $\phi_i$ for each component of the vector field
\begin{align}
\Psi_{l,\tau} = \phi_l \otimes \epsilon^{\tau}\,,
\end{align}
where $\epsilon^{\tau}$ is the basis of a vector.

On the other hand, the tensor fields $h_{ij}e^i\otimes e^j$ can be expanded using {\it ansatz} functions.
If the {\it ansatz} function is the same as the shape function, the scheme is called the Galerkin scheme. The FEM in this case is called the Galerkin FEM. Moreover, if the shape function is continuous, the method is also called the continuous Galerkin FEM. In this case, the tensor field $h_{ij}e^i\otimes e^j$ can be presented as 
\begin{align}
h_{ij}e^i\otimes e^j = H_{\sigma}\epsilon^{\sigma}=\tensor{H}{^k_{\sigma}}\phi_k\otimes\epsilon^{\sigma}\,,
\end{align}
where since each component $H_{\sigma}$ itself is a scalar field, it can be further expanded by the scalar test function $H_{\alpha}=\tensor{H}{^k_{\alpha}}\phi_k$.  

As Eq.~(\ref{weakwaveequation}) holds for any test functions $\Psi$, we can choose $\Psi$ as shape functions over all the different nodes in the domain. The left-hand-side of Eq.~(\ref{weakwaveequation}) then gives,
\begin{align}
\langle \phi_l \otimes \epsilon^{\tau},\frac{\partial^2}{\partial t^2}h_{ij} e^i\otimes e^j\rangle&=\langle \phi_l \otimes \epsilon^{\tau},\phi_k\otimes\epsilon^{\sigma}\rangle \frac{\partial^2}{\partial t^2}\tensor{H}{^k_{\sigma}}\nonumber\\
&=\langle \phi_l,\phi_k \rangle \otimes \tensor{\delta}{_{\tau}^{\sigma}}\frac{\partial^2}{\partial t^2}\tensor{H}{^k_{\sigma}}\,,
\end{align}
where the index $l$ runs over $1,...,N$ and $\tau$ runs over $1,...,5$.
Similarly, for the first term on the right-hand-side of Eq.~(\ref{weakwaveequation}), we obtain
\begin{align}
&\langle\phi_l \otimes \epsilon^{\tau}, c^2\nabla^2 h_{ij} e^i\otimes e^j \rangle=\langle\phi_l\otimes\epsilon^{\tau},c^2\nabla^2\phi_k\otimes\epsilon^{\sigma}\rangle\tensor{H}{^k_{\sigma}}\nonumber\\
=&-\langle(\nabla c^2)\phi_l,\nabla\phi_k\rangle \otimes\tensor{\delta}{_\tau^\sigma}\tensor{H}{^k_{\sigma}}-\langle c^2\nabla \phi_l,\nabla\phi_k\rangle \otimes\tensor{\delta}{_\tau^\sigma}\tensor{H}{^k_{\sigma}}\nonumber\\
&+\langle \phi_l, c^2\hat{n}\cdot \nabla \tensor{H}{_{\sigma}}\rangle_{\partial \Omega}\otimes\tensor{\delta}{_\tau^\sigma}\label{boundaryterms}
\end{align}
where for the last term, we have used integration by parts. $\langle \phi_l, c^2\hat{n}\cdot \nabla \tensor{H}{_{\sigma}}\rangle_{\partial \Omega}\otimes\tensor{\delta}{_\tau^\sigma}$ represents the integration over boundaries of the simulation domain $\partial \Omega$. This term is related to the boundary conditions in FEM, which we shall discuss in detail in the next few sections.

Similarly, for other terms, we have
\begin{align}
&\langle \phi_l\otimes\epsilon^{\tau},N^2\partial_i \Gamma_j e^i\otimes e^j\rangle\nonumber\\
=&\langle \phi_l\otimes\epsilon^{\tau},\left(f_{\kappa} \frac{\rho^m \rho^k}{\rho^2} \delta_{ki} h_{mj} +f_{\Gamma} \frac{\rho^m}{\rho} \partial_i h_{mj}\right) e^i\otimes e^j\rangle\nonumber\\
=&[\langle \phi_l,f_{\kappa}\frac{\rho^m \rho^k}{\rho^2} \delta_{ki} \tensor{C}{_m_j^\sigma}\tensor{C}{^i^j_\alpha}\phi_k\rangle\nonumber\\
&+\langle \phi_l,f_{\Gamma} \frac{\rho^m}{\rho}\tensor{C}{_m_j^\sigma}\tensor{C}{^i^j_\alpha}\partial_i \phi_k\rangle] \otimes\tensor{\delta}{_\tau^\alpha}\tensor{H}{^k_{\sigma}} \quad,\\
\nonumber\\
&\langle \phi_l\otimes\epsilon^{\tau},-2N^2\tensor{\Gamma}{^m_i_j}\Gamma_m e^i\otimes e^j\rangle\nonumber\\
=&-2\langle \phi_l,f_{\Gamma}\frac{\rho^l}{\rho}\tensor{C}{_l_m^\sigma} \tensor{\Gamma}{^m_i_j}\tensor{C}{^i^j_\alpha} \phi_k\rangle \otimes \tensor{\delta}{_\tau^\alpha} \tensor{H}{^k_{\sigma}}\quad,\\
\nonumber\\
&\langle \phi_l\otimes\epsilon^{\tau}, f_{\rho}\frac{\rho^m}{\rho}\partial_{m} h_{ij}e^i\otimes e^j\rangle\nonumber\\
=&\langle\phi_l\otimes\epsilon^{\tau},f_{\rho}\frac{\rho^m}{\rho}\partial_{m}\phi_k\otimes\epsilon^{\sigma}
\rangle\tensor{H}{^k_{\sigma}}\nonumber\\
=&\langle\phi_l,f_{\rho}\partial_{\rho}\phi_k\rangle\otimes\tensor{\delta}{_\tau^\sigma}\tensor{H}{^k_{\sigma}}\quad,\\
\nonumber\\
&\langle \phi_l\otimes\epsilon^{\tau},
f_R\delta^{mp}\tilde{R}_{ip}h_{mj} e^i\otimes e^j\rangle\nonumber\\
=& \langle \phi_l\otimes\epsilon^{\tau},f_R \delta^{mp}\tilde{R}_{ip}\tensor{C}{_m_j^{\sigma}}\tensor{C}{^i^j_{\alpha}}\phi_k\otimes\epsilon^{\alpha}\rangle\tensor{H}{^k_{\sigma}}\nonumber\\
=&\langle \phi_l,f_R \delta^{mp}\tilde{R}_{ip}\tensor{C}{_m_j^{\sigma}}\tensor{C}{^i^j_{\alpha}}\phi_k\rangle\otimes\tensor{\delta}{_\tau^\alpha}\tensor{H}{^k_{\sigma}}\quad,\\
\nonumber\\
&\langle\phi_l\otimes\epsilon^{\tau},f_R\delta^{mp}\delta^{nq}\tensor{\tilde{R}}{_p_i_j_q}h_{mn}e^i\otimes e^j\rangle\nonumber\\
=&\langle\phi_l\otimes\epsilon^{\tau},f_R\delta^{mp}\delta^{nq}\tensor{\tilde{R}}{_p_i_j_q}\tensor{C}{_m_n^{\sigma}}\tensor{C}{^i^j_{\alpha}}\phi_k\otimes\epsilon^{\alpha}\rangle\tensor{H}{^k_{\sigma}}\nonumber\\
=&\langle\phi_l,f_R\delta^{mp}\delta^{nq}\tensor{\tilde{R}}{_p_i_j_q}\tensor{C}{_m_n^{\sigma}}\tensor{C}{^i^j_{\alpha}}\phi_k\rangle\otimes\tensor{\delta}{_\tau^\alpha}\tensor{H}{^k_{\sigma}}\quad,\\
\nonumber\\
&\langle\phi_l\otimes\epsilon^{\tau},f_{\Gamma}\frac{\rho^p}{\rho}\partial_i h_{pj} e^i\otimes e^j\rangle\nonumber\\
=&\langle\phi_l\otimes\epsilon^{\tau},f_{\Gamma}\frac{\rho^p}{\rho}\tensor{C}{_p_j^{\sigma}}\tensor{C}{^i^j_{\alpha}}\partial_{i}\phi_k\otimes\epsilon^{\alpha}\rangle\tensor{H}{^k_{\sigma}}\nonumber\\
=&
\langle\phi_l,f_{\Gamma}\frac{\rho^p}{\rho}\partial_{i}\phi_k\tensor{C}{_p_j^{\sigma}}\tensor{C}{^i^j_{\alpha}}\rangle\otimes\tensor{\delta}{_\tau^\alpha}\tensor{H}{^k_{\sigma}}\quad,\\
\nonumber\\
&\langle \phi_l\otimes\epsilon^{\tau},f_{\Gamma}\frac{\rho^p}{\rho}\tensor{\Gamma}{^m_i_j} h_{pm} e^i\otimes e^j\rangle\nonumber\\
=&\langle\phi_l\otimes\epsilon^{\tau},f_{\Gamma}\frac{\rho^p}{\rho}\tensor{\Gamma}{^m_i_j}\tensor{C}{_p_m^{\sigma}}\tensor{C}{^i^j_{\alpha}}\phi_k\otimes\epsilon^{\alpha}\rangle\tensor{H}{^k_{\sigma}}\nonumber\\
=&
\langle\phi_l,f_{\Gamma}\frac{\rho^p}{\rho}\tensor{\Gamma}{^m_i_j}\tensor{C}{_p_m^{\sigma}}\tensor{C}{^i^j_{\alpha}}\phi_k
\rangle\otimes\tensor{\delta}{_\tau^\alpha}\tensor{H}{^k_{\sigma}}\quad.
\end{align}
Inserting the above expressions back into Eq.~(\ref{weakwaveequation}), we obtain $5\times N$ different equations, which form a linear system. In this case, it is more convenient  to present Eq.~(\ref{weakwaveequation}) in a matrix format
\begin{align}
\left\{ \begin{aligned}
\frac{\partial}{\partial t}\mathcal{H}&=\mathcal{V}\,\\
\mathcal{M}\frac{\partial}{\partial t}\mathcal{V} &=-\mathcal{M}^B\frac{\partial}{\partial t}\mathcal{H}-\mathcal{F}\mathcal{H}
\end{aligned}\,,
\right.
\end{align}
where $\mathcal{F}$ is defined by
\begin{align}
\mathcal{F}&=\mathcal{A}+\mathcal{D}^c-\mathcal{D}^{\rho}+2\mathcal{D}^{\rm Ricci}+2\mathcal{D}^{\rm Riemann}+2\mathcal{D}^{\kappa}\nonumber\\
&+4\mathcal{D}^{\Gamma_1}-4\mathcal{D}^{\rm \Gamma_2}\,.
\end{align}
The elements of matrices are defined by
\begin{equation}
\left\{ \begin{aligned}
\mathcal{M}_{(lk)\otimes(\tau\sigma)}&=\langle\phi_l,\phi_k\rangle\otimes \tensor{\delta}{_{\tau}^{\sigma}}\\
\mathcal{M}_{(lk)\otimes(\tau\sigma)}^B&=\langle c\phi_l,\phi_k
\rangle_{\partial \Omega}\otimes \tensor{\delta}{_{\tau}^{\sigma}}\\
\mathcal{D}_{(lk)\otimes(\tau\sigma)}^c&=\langle\nabla(c^2)\phi_l,\nabla\phi_k\rangle\otimes \tensor{\delta}{_{\tau}^{\sigma}}\\
\mathcal{A}_{(lk)\otimes(\tau\sigma)}&=\langle c^2\nabla\phi_l,\nabla\phi_k\rangle\otimes \tensor{\delta}{_{\tau}^{\sigma}}\\
\mathcal{D}_{(lk)\otimes(\tau\sigma)}^{\rho}&=\langle\phi_l,f_{\rho}\frac{\rho^m}{\rho}\partial_{m}\phi_k\rangle\otimes \tensor{\delta}{_{\tau}^{\sigma}}\\
\mathcal{D}_{(lk)\otimes(\tau\sigma)}^{\rm Ricci}&=\langle\phi_l,f_R \delta^{mp}\tilde{R}_{ip}\tensor{C}{_m_j^{\sigma}}\tensor{C}{^i^j_{\alpha}}\phi_k\rangle\otimes \tensor{\delta}{_\tau^\alpha}\\
\mathcal{D}_{(lk)\otimes(\tau\sigma)}^{\rm Riemann}&=\langle\phi_l,f_R\delta^{mp}\delta^{nq}\tensor{\tilde{R}}{_p_i_j_q}\tensor{C}{_m_n^{\sigma}}\tensor{C}{^i^j_{\alpha}}\phi_k\rangle\otimes \tensor{\delta}{_\tau^\alpha}\nonumber\\
\mathcal{D}_{(lk)\otimes(\tau\sigma)}^{\Gamma_1}&=\langle\phi_l,f_{\Gamma}\frac{\rho^p}{\rho}\tensor{C}{_p_j^{\sigma}}\tensor{C}{^i^j_{\alpha}}\partial_{i}\phi_k\rangle\otimes \tensor{\delta}{_\tau^\alpha}\\
\mathcal{D}_{(lk)\otimes(\tau\sigma)}^{\Gamma_2}&=\langle\phi_l,f_{\Gamma}\frac{\rho^p}{\rho}\tensor{\Gamma}{^m_i_j}\tensor{C}{_p_m^{\sigma}}\tensor{C}{^i^j_{\alpha}}\phi_k\rangle\otimes\tensor{\delta}{_\tau^\alpha}\\
\mathcal{D}_{(lk)\otimes(\tau\sigma)}^{\rm \kappa}&=\langle \phi_l,f_{\kappa}\frac{\rho^m \rho^k}{\rho^2} \delta_{ki} \tensor{C}{_m_j^\sigma}\tensor{C}{^i^j_\alpha}\phi_k\rangle\otimes\tensor{\delta}{_\tau^\alpha}\nonumber 
\end{aligned}\,.
\right. 
\end{equation}

\subsection{time discretization}
For the time discretization, we use the following scheme

\begin{align}
\mathcal{M}\frac{\mathcal{V}^n-\mathcal{V}^{n-1}}{k} =&-\mathcal{M}^B\frac{\mathcal{H}^n-\mathcal{H}^{n-1}}{k}\nonumber\\
&-\mathcal{F}[\theta \mathcal{H}^n+(1-\theta)\mathcal{H}^{n-1}]\,,\label{V0}\\
\frac{\mathcal{H}^n-\mathcal{H}^{n-1}}{k} =& \theta \mathcal{V}^n+(1-\theta)\mathcal{V}^{n-1} \,.\label{H0}
\end{align}

The superscript $n$ indicates the number of a time step and $k=t_n-t_{n-1}$ is the length of the present time step. Using Eq.~(\ref{V0}) to eliminate $\mathcal{H}^n$ in Eq.~(\ref{H0}) and Eq.~(\ref{H0}) to eliminate $\mathcal{V}^n$ in Eq.~(\ref{V0}), we can present $\mathcal{H}^n$ and $\mathcal{V}^n$ in terms of $\mathcal{V}^{n-1}$ and $\mathcal{H}^{n-1}$ 

\begin{align}
 [&\mathcal{M}+k\theta \mathcal{M}^B+k^2\theta^2\mathcal{F}]\mathcal{H}^{n}\nonumber\\
=&[\mathcal{M}+k\theta \mathcal{M}^{B}-k^2\theta(1-\theta)\mathcal{F}]\mathcal{H}^{n-1}+k\mathcal{M}\mathcal{V}^{n-1}\label{eqH}
\end{align}

\begin{align}
&[\mathcal{M}+k\theta \mathcal{M}^B+k^2\theta^2\mathcal{F}]\mathcal{V}^{n}\nonumber\\
=&[\mathcal{M}-k(1-\theta) \mathcal{M}^{B}-k^2\theta(1-\theta)\mathcal{F}]\mathcal{V}^{n-1}-k\mathcal{F}\mathcal{H}^{n-1}
\label{eqV}\,.
\end{align}

Given the knowledge of $\mathcal{V}^{n-1}$ and $\mathcal{H}^{n-1}$ at a previous time step, we can solve $\mathcal{H}^{n}$ and $\mathcal{V}^{n}$ from the above two linear equations.

When $\theta = 0$, the scheme is called the forward or explicit Euler method. If $\theta = 1$, it reduces to the backward or implicit Euler method. The scheme adopted in this work is called the {\it Crank-Nicolson Scheme}, namely $\theta = 1/2$, which uses the midpoint between two different time steps. This scheme is {\it implicit} and is of second-order accuracy. An advantage of the {\it implicit} method is that it can be stable for arbitrary step sizes if the scheme is {\it upwind} (see. Chapter 2 in Ref.~\cite{grossmann2007numerical}). However, a stable scheme does not guarantee a correct solution to the wave equation. We need to resolve the waveforms as well. For this purpose, in our simulations, we set $k=\lambda/10$ and the size of the mesh $\sigma<\lambda/15$, where $\lambda$ is the wavelength.

\subsection{linear solvers}
The number of independent equations in Eqs.~(\ref{eqH}) and~(\ref{eqV}) is called the degree-of-freedom (DOF) of the system. In FEM, the DOF is usually very large, which can be easily up to $10^9$. Therefore, direct methods such as the LU decomposition are inefficient in this case. One has to use the iterative methods. However, it is important to note that, matrices $D^{c}\,$, $D^{\rho}\,$, $D^{\Gamma_1}\,$,$D^{\Gamma_2}\,$,$D^{\kappa}\,$,$D^{\rm Ricci}\,$,$D^{\rm Riemann}\,$are not symmetric in our case. Some conventional methods such as the Conjugate Gradient (CG) method can not be used here. Instead, we use the GMRES method (a generalized minimal residual algorithm for solving non-symmetric linear systems)~\cite{osti_409863}, which does not require any specific properties of the matrices. 

In practice, the efficiency of the GMRES method is also strongly dependent on the preconditioners used. A good preconditioner can significantly reduce the number of iterations needed in the GMRES method. Moreover, for massively paralleled linear solvers, the primary bottleneck, indeed, comes from the difficulty to produce preconditioners that can scale up to a large number of processors, rather than from the communication between processors. Fortunately, it has been shown in the past decade that the algebraic multigrid (AMG) method~\cite{brandt1982algebraic}, which can be used to construct preconditioner only based on the matrix itself, is suitable for massive parallelization and is extremely efficient in this case. Therefore, in this work, we adopt the GMRES linear solver together with the AMG preconditioner. As a result, in our method, it only takes less than $25$ iterations for the linear system to achieve a rather stringent convergence criterion 
\begin{align}
\left\|\rm Residual \right\|_{\infty,h}:= \max_{x_h\in \overline{\Omega}_h}| {\rm Residual}(x_h)|<10^{-14}\,,
\end{align}
where $\overline{\Omega}_h$ denotes all the grid points inside the domain and at its boundary $\overline{\Omega}_h=\Omega_h+\partial \Omega_h$.

\subsection{Boundary conditions}
For test functions $\phi \in \mathbb{V}$, the Dirichlet boundary conditions do not appear explicitly in Eqs~(\ref{eqH},\ref{eqV}), which are called {\it essential boundary conditions}. However, the Neumann conditions have to appear explicitly in the formulation, which is called {\it natural boundary conditions}. The boundary conditions only appear in these terms associated with the Laplacian operator $\nabla^2$, namely, the last term in Eq.~(\ref{boundaryterms})
\begin{align}
&\langle \phi_l, c^2\hat{n}\cdot \nabla \tensor{H}{_{\sigma}}\rangle_{\partial \Omega}\otimes\tensor{\delta}{_\tau^\sigma}\nonumber\\
=&\langle \phi_l, c^2\hat{n}\cdot \nabla \tensor{H}{_{\sigma}}\rangle_{\partial \Omega_1}\otimes\tensor{\delta}{_\tau^\sigma}+\langle \phi_l, c^2\hat{n}\cdot \nabla \tensor{H}{_{\sigma}}\rangle_{\partial \Omega_2}\otimes\tensor{\delta}{_\tau^\sigma}\,,
\end{align}
where $\partial\Omega_1$ represents the boundaries of the simulation domain and $\partial\Omega_2$
represents the horizon of black hole. 

On $\partial\Omega_1$ we impose the absorbing boundary condition
\begin{align}
\langle \phi_l, c^2\hat{n}\cdot \nabla \tensor{H}{_{\sigma}}\rangle_{\partial \Omega_1}\otimes\tensor{\delta}{_\tau^\sigma}=-\langle c\phi_l,\phi_k\rangle_{\partial \Omega_1}\otimes\tensor{\delta}{_\tau^\sigma}\frac{\partial}{\partial t}\tensor{H}{^k_{\sigma}}\,,
\end{align}
where we have used
\begin{equation}
\hat{n}\cdot\nabla \tensor{H}{_{\sigma}}=-\frac{1}{c}\frac{\partial{\tensor{H}{_{\sigma}}}}{\partial{t}}\quad {\rm on} \quad \partial\Omega_1\times (0,T] \,.\label{boundary2}
\end{equation}
The absorbing boundary condition is also called the {\it non-reflecting boundary} conditions or {\it radiation boundary} condition. These boundary conditions can eliminate the reflections of waves on boundaries, which enables us to simulate the propagation of waves in free space using a finite simulation domain.

At the horizon $\partial \Omega_2$, we have
\begin{equation}
\lim_{\rho \rightarrow M/2 }c^2=0\,.
\end{equation}
A notable feature of our formulation is that
the boundary term on $\partial \Omega_2$ vanishes 
\begin{equation}
\langle \phi_l, c^2\hat{n}\cdot \nabla \tensor{H}{_{\sigma}}\rangle_{\partial \Omega_2}\otimes\tensor{\delta}{_\tau^\sigma} = 0\,.
\end{equation}
As such, our formalism can naturally avoid artificial boundary conditions at the horizon.

\subsection{Numerical implement}
The numerical implement of this work is based on our code {\bf GWSIM}~\cite{He:2021hhl}, which is further based on the public available code {\bf deal.ii}~\cite{dealII91,BangerthHartmannKanschat2007,dealII90}. {\bf deal.ii} is a C++ program library, which is designed to solve partial differential equations based on modern finite element method. Coupled to efficient stand-alone linear algebra libraries, such as PETSc~\cite{abhyankar2018petsc,petsc-web-page,petsc-user-ref,petsc-efficient},  {\bf deal.ii} supports massively parallel computing of large sparse linear systems of equations. {\bf deal.ii} also provides convenient tools for {\it triangulation} of various geometries of the simulation domain. 

\section{Numerical results \label{sec::numres}}
\begin{figure*}[!htb]
{\includegraphics[width=\linewidth]{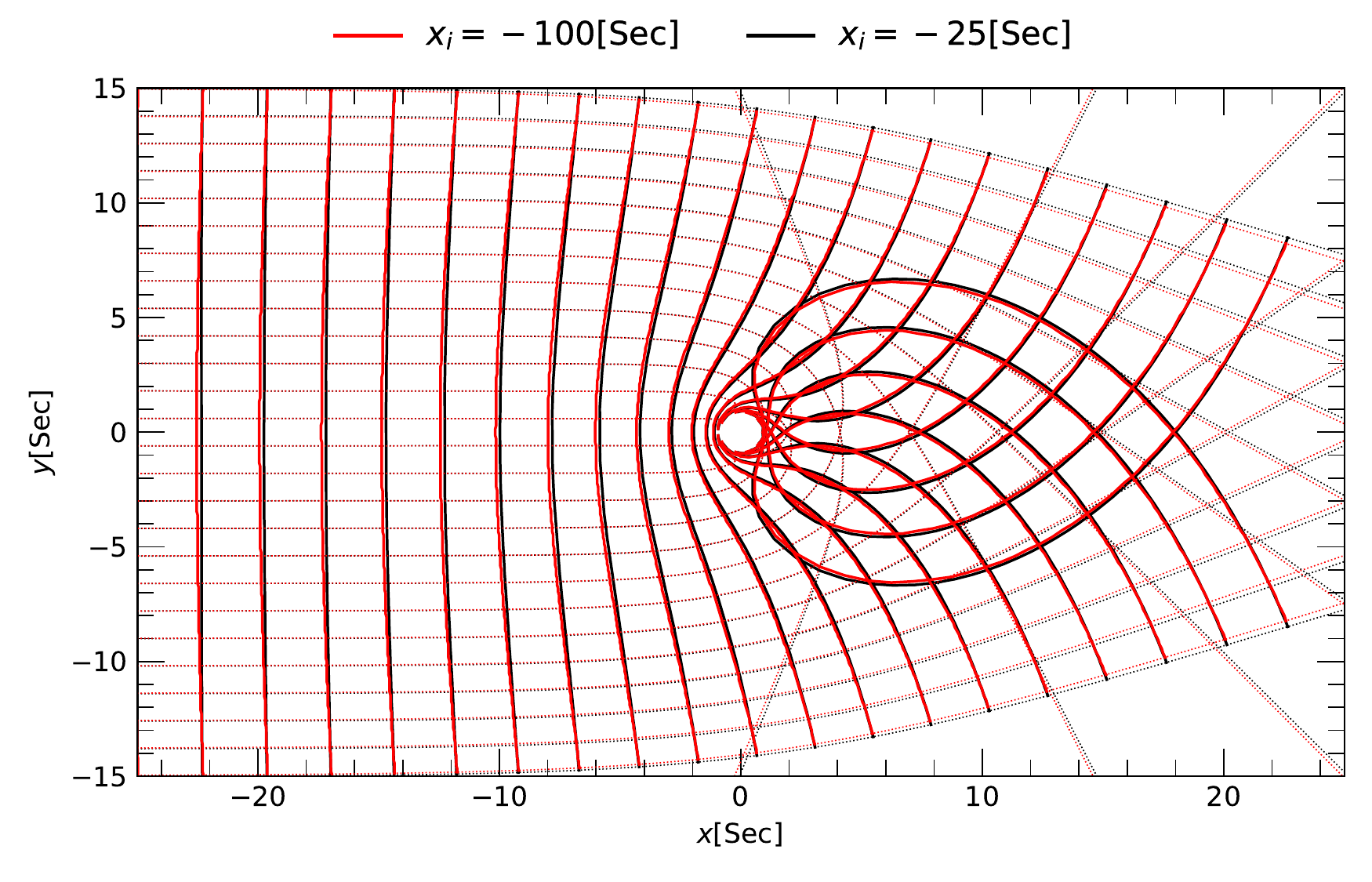}}
\caption{The evolution of the wavefronts of GWs in Schwarzschild spacetime predicted using null geodesics. The dashed lines are for congruence of null geodesics, which are also integral curves of the normal vectors of the wavefront. The solid lines show wavefronts at different times. The mass of the black hole in our code is $M=3\times 10^5 M_{\odot}$. Red and black colors show the null geodesics obtained by setting initial conditions at $x_i=-100\,[{\rm Sec}]$ and $x_i=-25\,[{\rm Sec}]$, respectively. If the distance from the center of the black hole is above $25\,[{\rm Sec}]$, the effect of distortion on the initial wavefront due to the black hole itself is negligible in the regimes of our simulations. Moreover, above $25\,[{\rm Sec}]$ there are also no appreciable differences in setting initial conditions at different places on the wavefront. Note that due to the linearity of the wave equation and the geometric unit, the parameters used in our code and numerical simulations can be rescaled to other values that are of astrophysical interest.
\label{wavefronttheory}}
\end{figure*}
\subsection{Null geodesics and wavefronts}
The basic setups follow our previous work~\cite{He:2021hhl}. We assume that the source of GWs is far away from the simulation domain and the incident waves travel along the axis of the cylinder ($x$-axis). However, due to the long-range nature of gravity produced by the black hole, the incident waves suffer the Shapiro time delay on their way from the source to the black hole, which adds up a shift of the arrival time to the wavefronts relative to the case without such a black hole and may also distort the shape of the wavefront that arrives at the black hole. However, as pointed out in our previous work~\cite{He:2021hhl}, these effects can be accurately and robustly determined using null geodesics as tracers for the wavefronts of GWs. This is because from Eq.~(\ref{wavespeed_light_rays}), GWs travel at the same speed as those of light rays in Schwarzschild spacetime. The wave vector of GWs, therefore, is a null vector $k^ak_a=0$. As a result, by definition the hypersurface of the wavefront is a null hypersurface. A key property of null hypersurface is that the integral curves of its normal vector $k^a$ are null geodesics. This holds even in the curved spacetime (see equation 4.2.37 in~\cite{wald2010general}). Therefore, the wavefront of GWs in Schwarzschild spacetime can be accurately traced by null geodesics. The equations for null geodesics in Schwarzschild spacetime in isotropic coordinates are provided in Appendix~\ref{Nullgeodesic}.
After obtaining the spatial trajectories of null geodesics, the Killing time of each wavefront can be obtained by integrating
\begin{align}
dt=\frac{dl}{c}\,,
\end{align}
along the geodesics, where $dl$ is the spatial length and $c$ is the isotropic wave speed.

\begin{figure}
{\includegraphics[width=\linewidth]{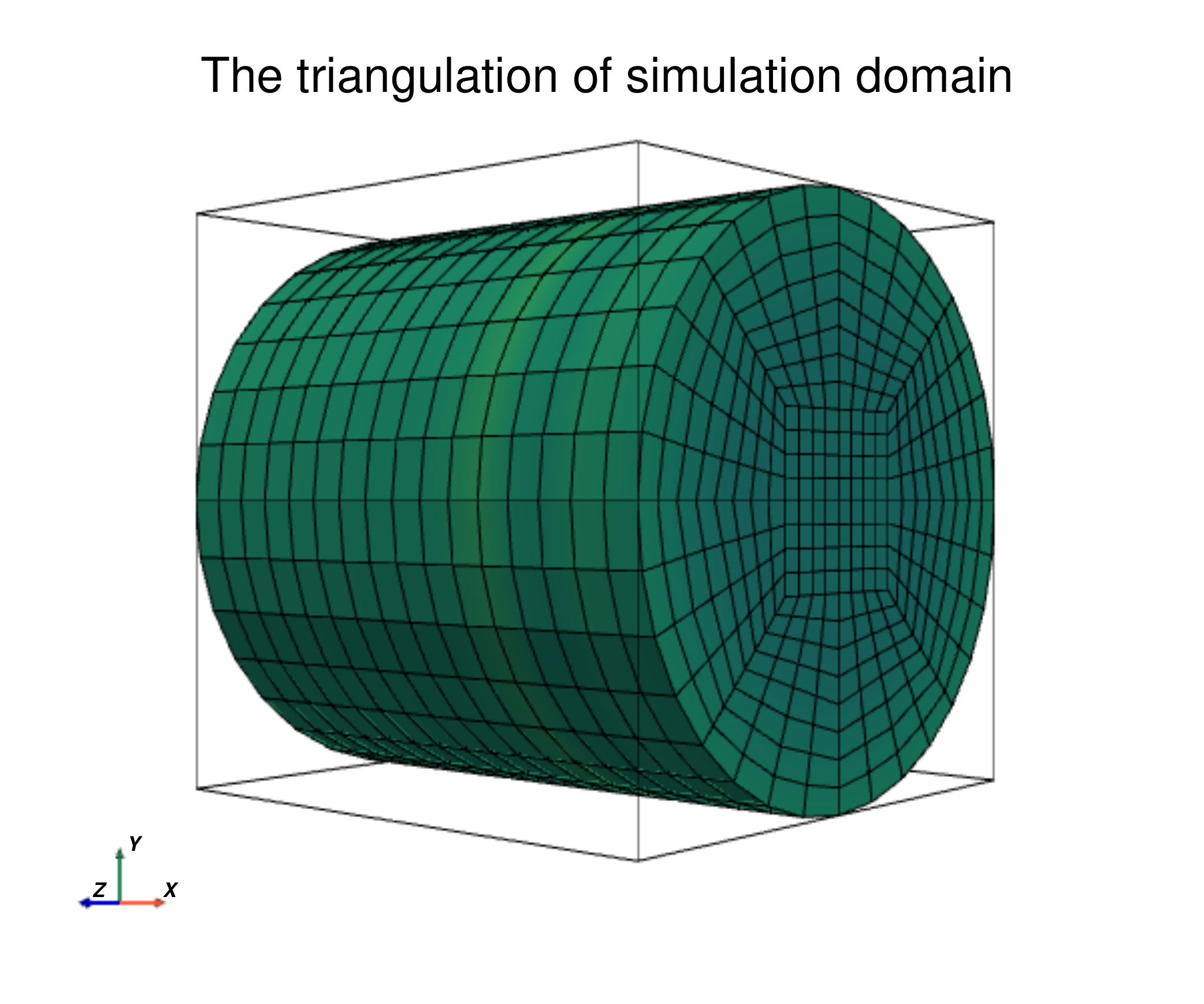}}
\caption{The triangulation of our simulation
domain. We only show the refinement of $2^5$ (along one dimension ) for illustrative purposes. The incident waves travel along the $x$-axis from $x=-\infty$ to $x= +\infty$. The wavefront of the incident wave at $x=-L_{\rm cylinder}/2$ is normal to the $x$-axis.
\label{cylinderical}}
\end{figure}

Figure~\ref{wavefronttheory} shows the evolution of the wavefronts of GWs in Schwarzschild spacetime. The dashed lines are for null geodesics, which are also the integral curves of the normal vectors of the wavefront. The solid lines show wavefronts at different times. 
In our code, the mass of black hole is $M=3\times 10^5 M_{\odot}$. Red and black colors show the null geodesics obtained with initial conditions at $x_i=-100\,[{\rm Sec}]$ and $x_i=-25\,[{\rm Sec}]$, respectively. If the distance to the center of black hole is above $25\,[{\rm Sec}]$, the distortion effect of the black hole on the initial wavefront is negligible in the regimes of our simulations. Moreover, above $25\,[{\rm Sec}]$ there is also no appreciable impact of setting initial conditions at different places on the wavefronts.  

Given the above tests, we choose the simulation domain as a cylinder with a length of $L_{\rm cylinder}=50\,[{\rm Sec}]$. In this case, the distance from the boundary at $x=-L_{\rm cylinder}/2$ to the black hole is long enough so that the wavefront of the incident waves is not distorted by the black hole at the boundary of the simulations domain. As such, the incident GWs at $x=-L_{\rm cylinder}/2$ can be considered as plane waves with a constant wave vector $\vec{k}$ along the $x$-axis. Figure~\ref{cylinderical} shows the triangulation of our simulation
domain. For illustrative purposes, we only show the refinement of $2^5$ (along one dimension). Unlike a cubic domain which is usually used in numerical simulations, a cylinder domain can minimize the impact of boundaries on waveforms as both the simulation domain and  wavefront are of the azimuth symmetry. 

Further note that due to the linearity of the wave equation and the geometric unit, the parameters used in our code and numerical simulations can be rescaled to other values that are of astrophysical interest. 

\subsection{Wave zones}
As pointed out in the previous section, the leading  wavefront can be traced by null geodesics, which provides a robust test of our simulation results. 

In this test, we choose the radius of the cylinder as 
$R_{\rm cylinder}=15\,[{\rm Sec}]$. The incident waves are set at $x=-L_{\rm cylinder}/2$ as
\begin{equation}
\left\{
\begin{aligned}
&H_0=h_{22}=h(t)\nonumber\\
&H_1=h_{\times}=0\nonumber\\
&H_2=0\nonumber\\
&H_3=0\nonumber\\
&H_4=0\nonumber
\end{aligned} \right.\,,
\end{equation}	
which are along the $x$ direction. In this test, we assume that the incident wave has only one non-vanishing polarization pattern. Given that $H_2=0$ initially, we have $h_{zz}=-H_2-h_{yy}=-h_{yy}$ and
$h_{+}=h_{yy}=-h_{zz}$ for the incident wave. Moreover, since the wave vector of the incident wave is along the $x$-axis $\vec{k}\parallel \vec{x}$, the incident wave is transverse relative to the $x$-axis.

We first choose the wave train as a wavelet. The waveform of the input wave train is a sinusoidal wave but only lasting for one period 
\begin{equation}
h(t)=\left\{
\begin{aligned}
A\sin(\omega t)\,, t\le\lambda\\
0\,,  t>\lambda 
\end{aligned} \right.\,, \label{sinusoidal}
\end{equation}
where $\omega = 2\pi/\lambda$ is the angular frequency and $\lambda = 5\,[{\rm Sec}]$ is the wavelength. The amplitude $A$ is normalized as unity. The black hole is placed at the center of our simulation domain with a mass of $M=3\times 10^5 M_{\odot}\,$, the same as before. The radius of the horizon of the black hole is $\rho_s=M/2=0.738803\,[{\rm Sec}]$. We perform a high-resolution simulation with a refinement of $2^8$. The total DOFs is $8.055\times 10^8$. The simulation uses $1920$ CPU cores and the total cost is $236.6{\rm K}$ CPU-hours. The parameters of our simulation are summarized in Table~\ref{Table_Re}.

\begin{table*}
\centering
\caption{The parameters used in our simulation \label{Table_Re}}
\begin{tabular}{c|c|c|c|c}
\hline 
Refinement & Total DOF & mass of black hole &  maximum diameter of elements & time step   \\ 
\hline 
$2^8$ & $8.055\times 10^8$ & $3\times 10^5 M_{\odot}$  & $0.109055[{\rm Sec}]$ & $0.00839504[{\rm Sec}] $ \\ 
\hline
\end{tabular} 
\end{table*}

\begin{figure*}[!htb]
{\includegraphics[width=\linewidth]{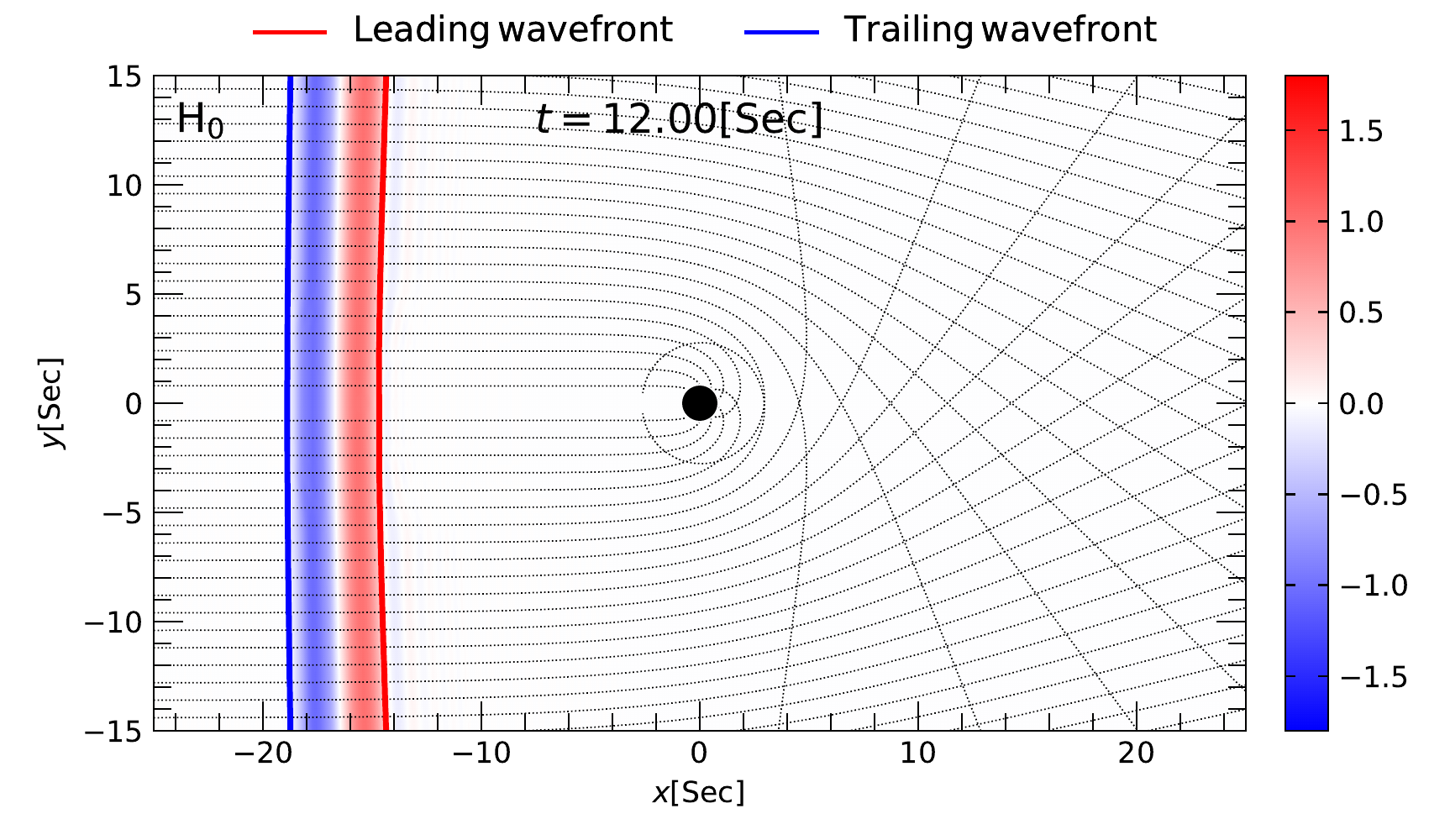}}
\caption{The wave train of $H_0$ in the Schwarzschild spacetime. The snapshot is taken at $t=12.00\,[{\rm Sec}]$ along the $x-y$ plane at $z=0$. The solid black circle indicates the position of the black hole with a radius of $\rho_s=0.738803\,[{\rm Sec}]$, the same as the horizon of the black hole. The dotted lines are for the congruence of null geodesics. The red and blue solid lines show the leading (red) and trailing (blue) wavefronts predicted by null geodesics, respectively. The color bar to the right shows the amplitude of the wave train in our simulations. 
\label{sim1}}
\end{figure*}

Figure~\ref{sim1} shows the wave train in the Schwarzschild spacetime. The snapshot is taken at $t=12.00\,[{\rm Sec}]$ along the $x-y$ plane at $z=0$. The solid black circle indicates the position of the black hole. The radius of the horizon of the black hole is $\rho_s=0.738803\,[{\rm Sec}]$. The dotted lines are for the congruence of null geodesics. The red and blue solid lines show the leading (red) and trailing (blue) wavefronts predicted by null geodesics, respectively. The color bar to the right shows the amplitude of the wave train. The wave zone in our simulations agrees with the theoretical predictions very well.   

\begin{figure*}[!htb]
{\includegraphics[width=\linewidth]{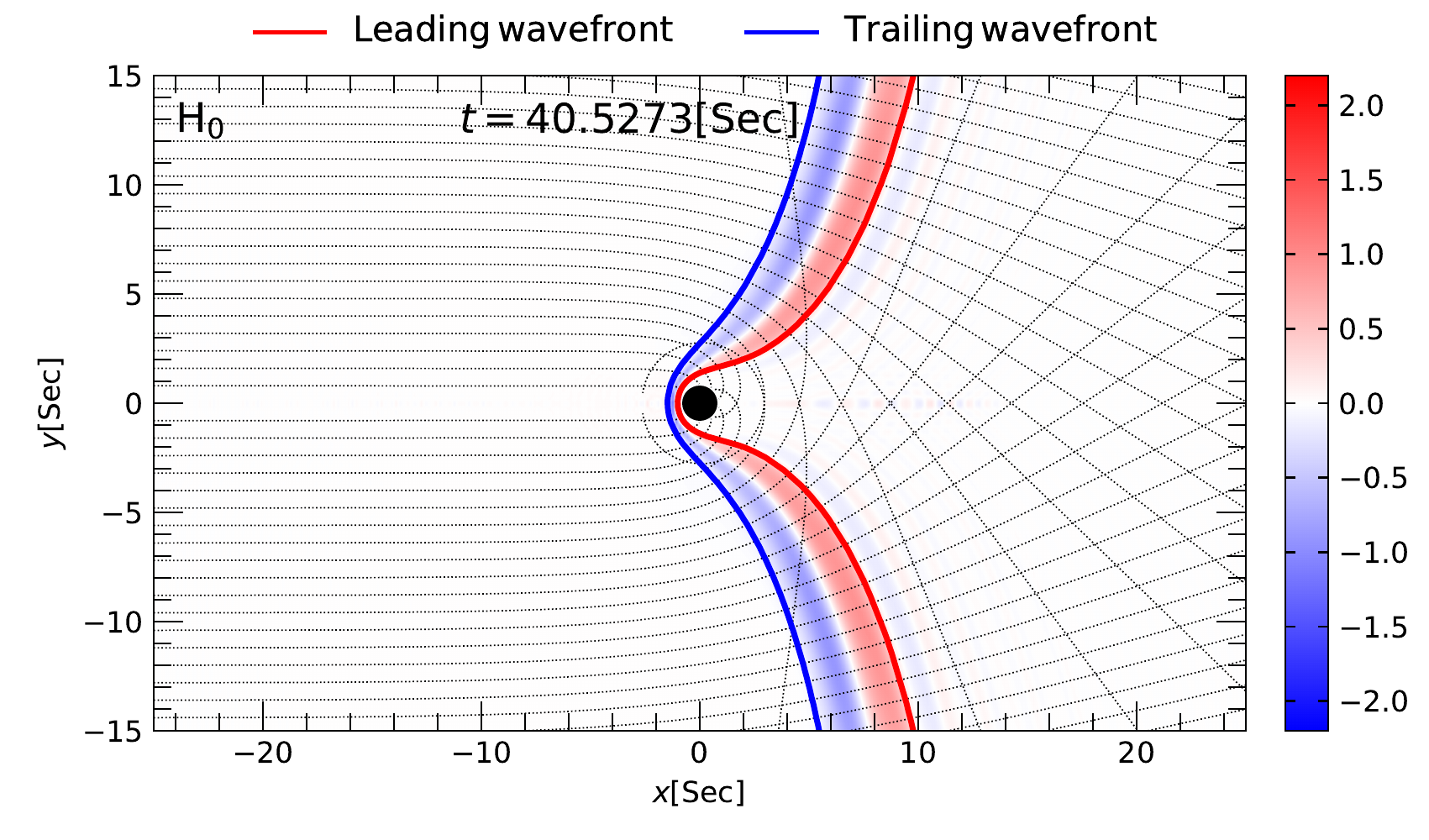}}
\caption{Similar to Fig.~\ref{sim1} but for the wave train at $t=40.5273\,[{\rm Sec}]$. At this instant, GWs have already arrived at the black hole. Due to the fact that GWs travel faster at outer regions than those close to the horizon, the wave zones are bent toward the black hole.  
\label{sim2}}
\end{figure*}

Figure~\ref{sim2} shows the wave train at $t=40.5273\,[{\rm Sec}]$. At this time, GWs have already arrived at the black hole. Due to the fact that GWs travel faster at outer regions than those close to the horizon, the wave zone is bent toward the black hole. 
\begin{figure*}[!htb]
{
\includegraphics[width=\linewidth]{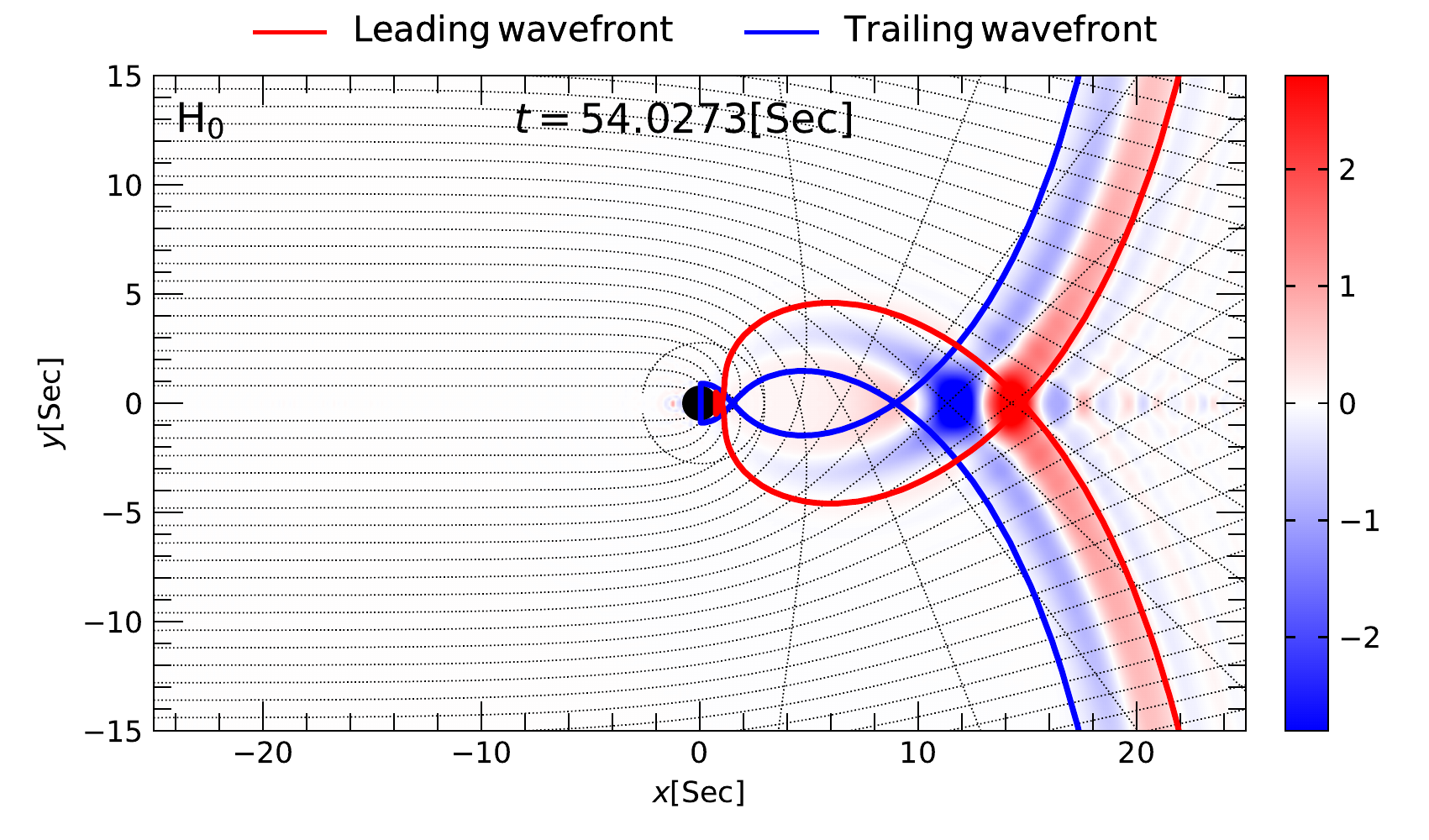}
}
\caption{Similar to Fig.~\ref{sim1} but for waves at $t=54.0273\,[{\rm Sec}]$. At this instant, most GWs have passed through the black hole. However, the wave zone of GWs is wildly twisted, which has a complicated geometry. Indeed, unlike spherical infall waves, most GWs in our case simply pass by and are deflected by the black bole. Only in a small region along the x-axis (radius), GWs can be reflected back when they hit the horizon. 
\label{sim3}}
\end{figure*}

Figure~\ref{sim3} shows the wave train at $t=54.0273\,[{\rm Sec}]$. At this time, GWs have already passed through the black hole. The wave zone is wildly twisted. Unlike in geometric optics, GWs do not form a caustic behind the black hole at scales comparable to their wavelength. Instead, an interference pattern appears in the overlaps of the twisted wave zones. Moreover, most GWs, indeed, only simply pass by and are deflected by the black bole. Only in a small region along the x-axis (radius), GWs can be reflected back when they hit the horizon. Moreover, ahead the leading wavefront, wavefront shocks appear along the $x$-axis. These wavefront shocks are numerical artifacts. The significance of such shocks is dependent on the relative resolution between the wavelength of the input waves and the mesh size of simulations. As shown in our previous work~\cite{He:2021hhl}, they can be significantly suppressed with a higher resolution in  simulations.

\begin{figure*}[!htb]
{
\includegraphics[width=\linewidth]{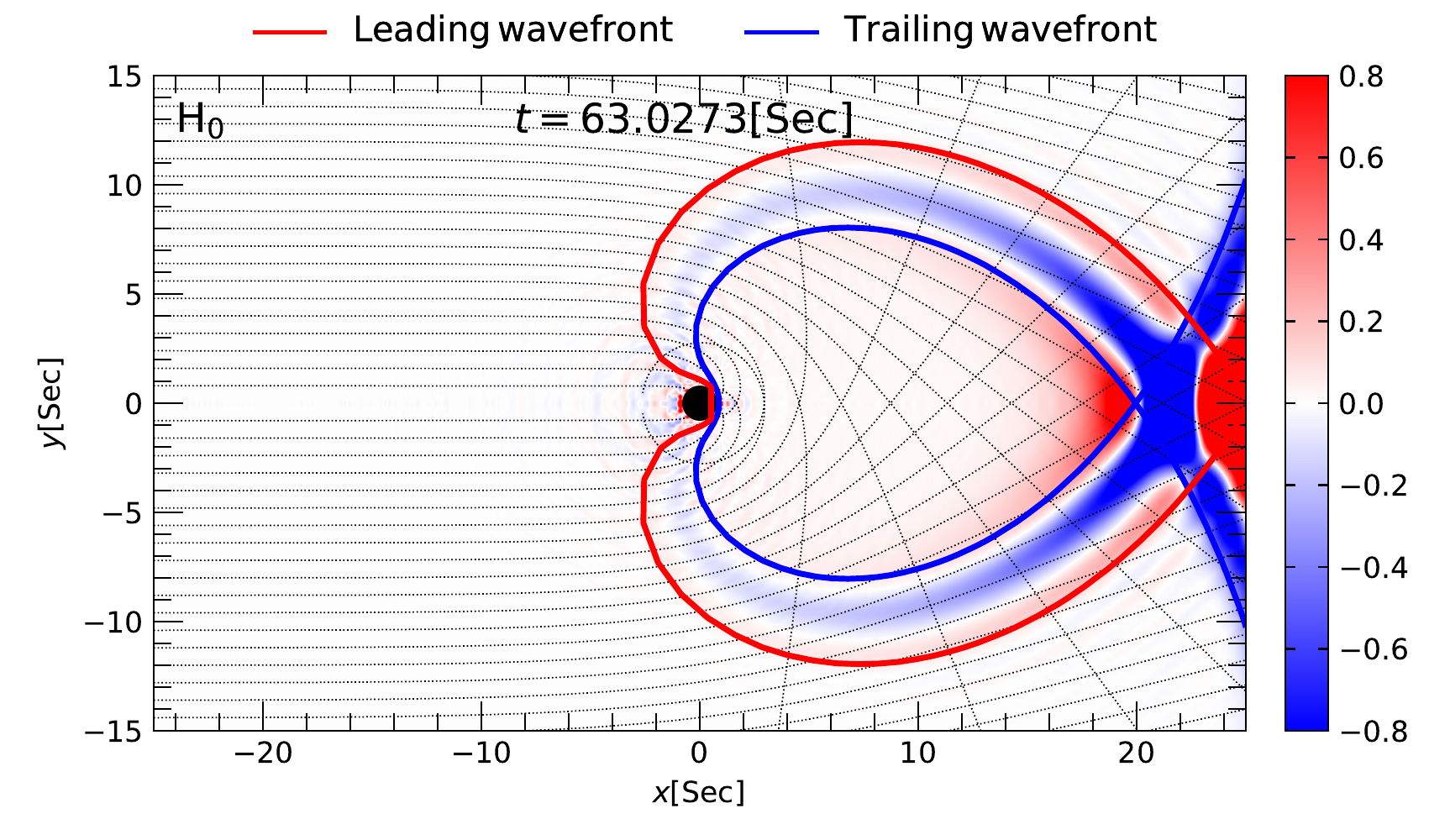}
}
\caption{Similar to Fig.~\ref{sim1} but for waves at $t=63.0273\,[{\rm Sec}]$. Unlike in the case of flat spacetime, the strong Huygens’ principle of waves does not hold in the curved spacetime due to the effect of back-scattering between GWs and the curvature of the spacetime. A tail behind the trailing wavefront (red shaded regions ) emerges. The significance of such effect is dependent on the direction of propagation of the trailing wavefront. If the trailing wavefront propagates perpendicularly to the radius of the black hole, there is no such back-scattering effect and the trailing wavefront is still clear. However, if the trailing wavefront propagates along the radius, a clear tail (red shaded regions) emerges.
\label{sim4}}
\end{figure*}

Figure~\ref{sim4} shows the wave train at $t=63.0273\,[{\rm Sec}]$. 
Unlike in the case of flat spacetime, the strong Huygens’ principle of waves does not hold in curved spacetime due to the effect of back-scattering. A tail behind the trailing wavefront (red shaded regions ) emerges. However, the significance of such an effect is dependent on the direction of propagation of the trailing wavefront. If the trailing wavefront travels perpendicularly to the radius of the black hole, there is no such back-scattering effect and the trailing wavefront is still clear. However, if the trailing wavefront travels along the radius, a clear tail emerges. Physically speaking, the effect of back-scattering is caused by the interaction between GWs and the background curvature. If GWs travel perpendicular to the radius of the black hole, the curvature of the background spacetime relative to GWs does not change. And, as a result, there are no such interactions and back-scattering. However, if GWs travel along the radius, curvature changes and such back-scattering emerges. 
\subsection{Power-law tail}
Besides the back-scattering, another intrinsic response of a black hole to a perturbation is called quasinormal modes (QNMs) (see~\cite{2009CQGra..26p3001B} for a review). However, QNMs happen only under particular boundary conditions, the energies of which blow up both at the horizon and spatial infinity. As a result, QNMs do not form a complete set of wavefunctions and it is in general impossible to represent a regular wavelet like ours as a sum of QNMs. Indeed, despite QNMs having been known for over three decades, how these modes are actually excited by physically relevant perturbations is less well studied~\cite{2009CQGra..26p3001B}.

\begin{figure}[!htb]
{
\includegraphics[width=\linewidth]{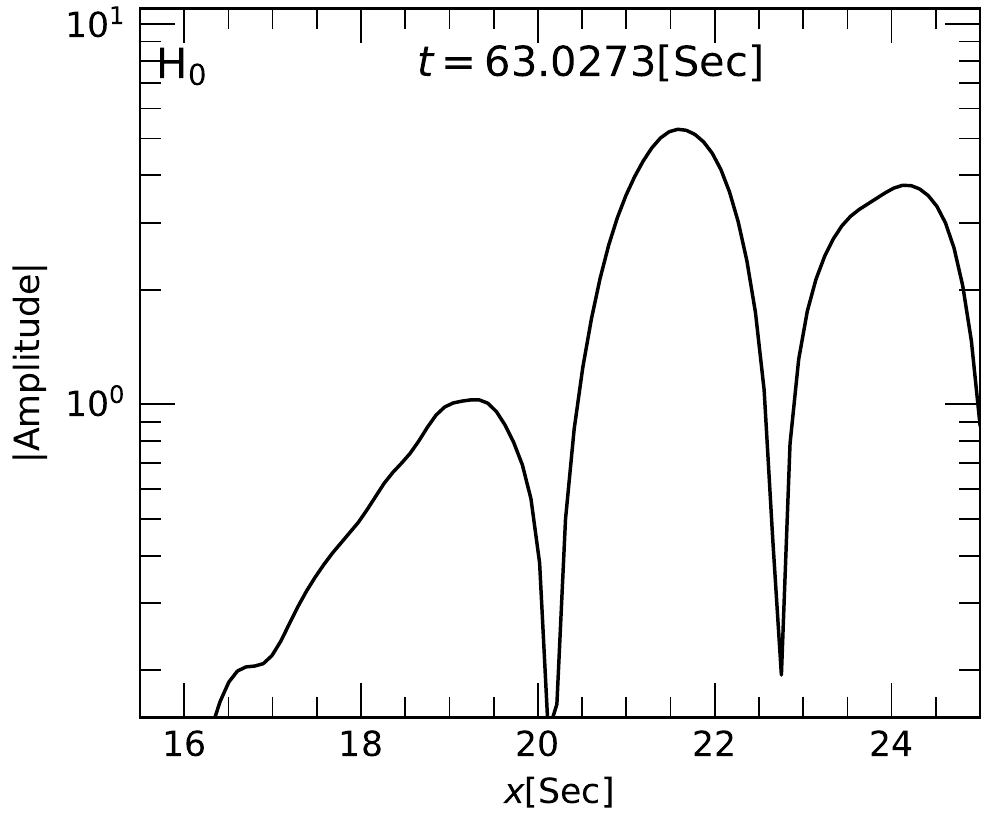}
}
\caption{The absolute value of the spatial waveform along the x-axis $(y=0,z=0)$ at  $t=63.0273\,[{\rm Sec}]$. A clear power-law tail appears behind the trailing wavefront.
\label{powerlawtail}}
\end{figure}

Our results suggest that QNMs if any, are sub-dominant in our case, as the amplitude of the incident wavelet on the x-axis is amplified behind the black hole, which is the dominant signal. However, as shown in Fig.~\ref{powerlawtail}, a power-law tail emerges behind the trailing wavefront, which is significant in our case. It is worth noting that such a tail does not appear before the black hole as in this case the trailing wavefront travel faster than the leading wavefront.

\subsection{Sinusoidal Wave}
Next, we choose the input GWs as a continuous sinusoidal wave, the same functional form as Eq.(\ref{sinusoidal}) but for a much longer time. The parameters of the simulation are the same as those in Table~\ref{Table_Re}.

\begin{figure*}[!htb]
{\includegraphics[width=\linewidth]{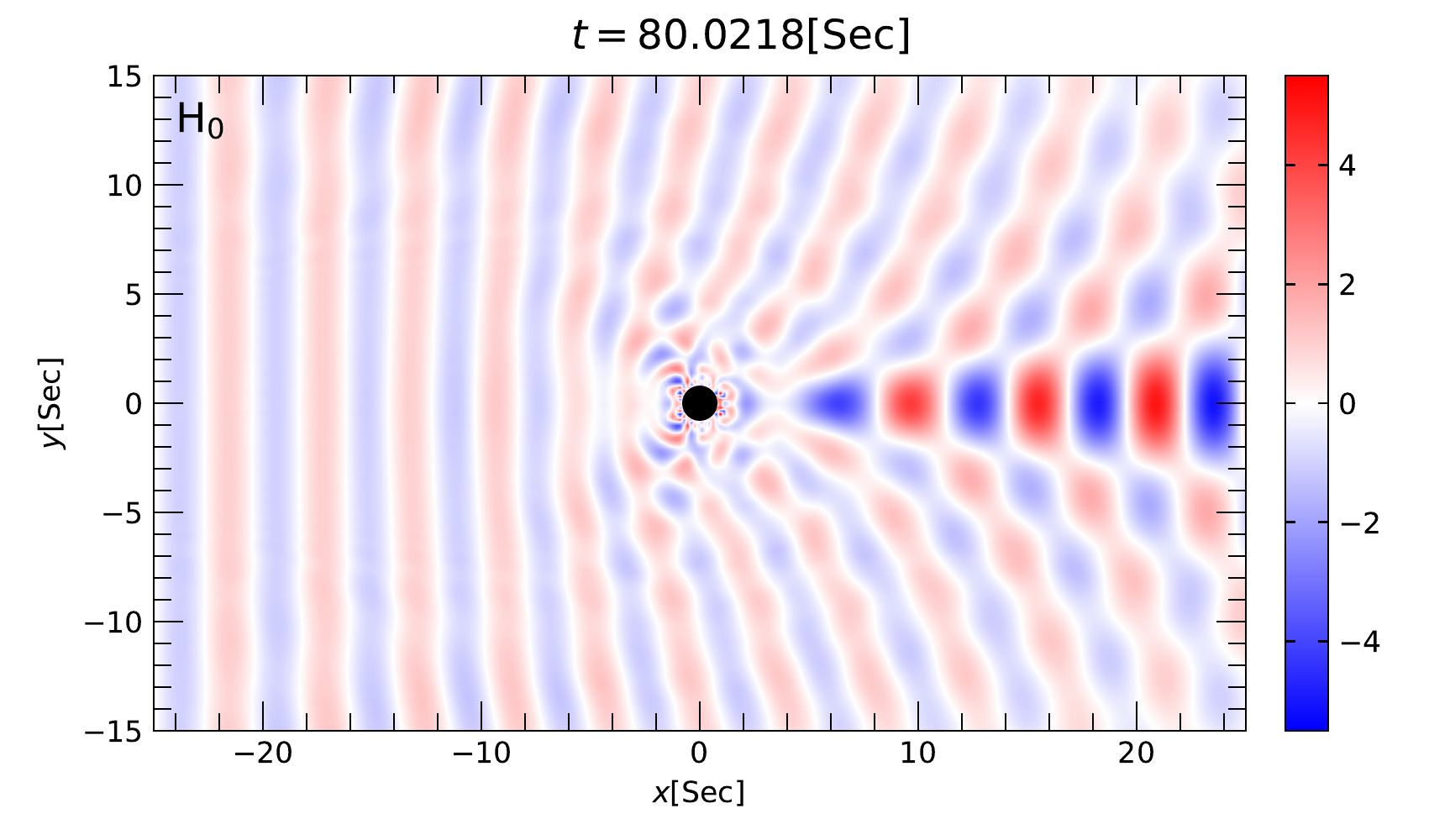}}
\caption{A snapshot of a continuous sinusoidal wave at $t=80.0218\,[{\rm Sec}]$ and $z=0$ along the $x-y$ plane. The waves are in a steady state. Unlike geometric optics, GW signals cannot be sheltered by the black hole due to the wave nature of GWs. GWs also do not form a caustic behind the black hole at scales that are comparable to their wavelength. Instead, a strong beam and an interference pattern appear in the overlaps of the wave zones along the optical axis behind the black hole. 
\label{sim5}}
\end{figure*}

Figure~\ref{sim5} shows a snapshot at $t=80.0218\,[{\rm Sec}]$ and $z=0$ along the $x-y$ plane. The waves are in a steady state. Unlike geometric optics, GW signals cannot be sheltered by the black hole due to the wave nature of GWs. GWs do not form a caustic behind the black hole at scales that are comparable to their wavelength. Instead, a strong beam and an interference pattern appear in the overlaps of the wave zones along the optical axis behind the black hole.

\subsection{Realistic input GW waveform}
 In this subsection, we choose the input waveform as a realistic template generated by binary black holes. The waveform is generated by NRHybSur3dq8 model~\cite{Varma:2018mmi}, which is a surrogate model for numerical relativity simulations. The initial total mass of the binary black holes is $40 M_{\odot}$. The binary has equal masses. We choose the lens as an isolated black hole with a mass of $133.33 M_{\odot}$. The binary GW source is $100.0 {\rm Mpc}$ away from the lens. The inclination of the angular momentum plane of the source binary is $\pi/2$ so that only $h_+$ polarization can be observed at the lens. We choose the starting time of the waveform at a point $0.094666$[Sec] before the merger of the binary black holes. 

\begin{figure*}[!htb]
{\includegraphics[width=\linewidth]{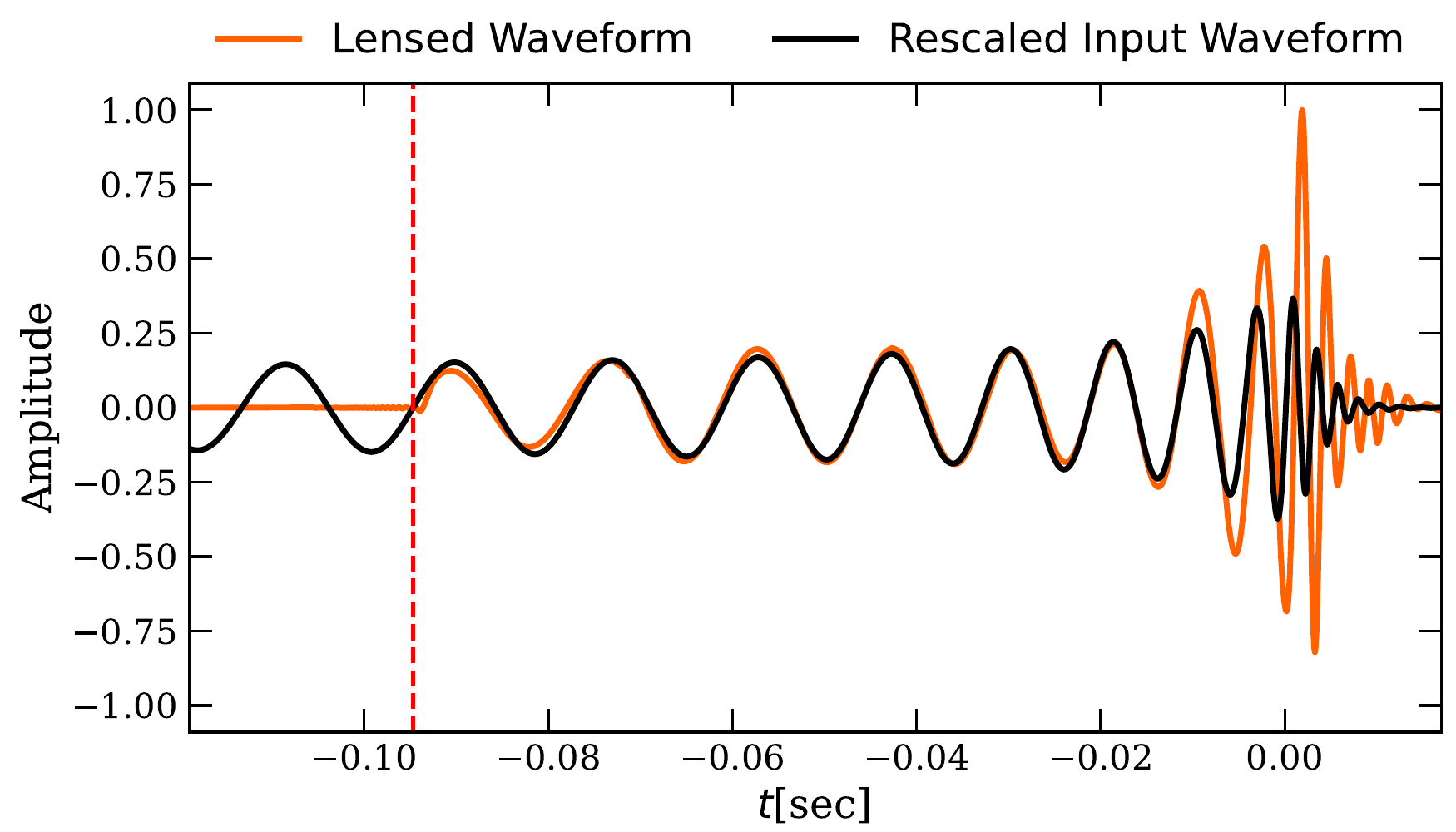}}
\caption{The input waveform is generated by binary black holes with equal masses (solid black curve). The initial total mass of the binary is chosen as $40 M_{\odot}$. The mass of the lens black hole is chosen as $133.33 M_{\odot}$. The starting time of the waveform in our simulation is at a point $0.094666$[Sec] before the merger of the binary black holes (the dashed vertical line). In practice, the simulation is performed with a scaling factor of $750$, in which the lens is $10^5 M_{\odot}$ in our simulation. The orange line shows the temporal lensed waveform after the black hole observed at $x=0.023333[{\rm Sec}]\approx 73\rho_s$ along the $x$-axis, which is far away from the horizon of the black hole. The amplitude of the lensed waveform is normalized by its maximum value. The initial amplitude of the input waveform is re-scaled to match the lensed waveform and its time is also shifted for the purpose of comparison. Because of the effect of back-scattering, the lensed waveform is much longer than that of the input waveform in the merger and ringdown phases. The relative strength of the lensed waveform in the merger and ringdown phases also exhibits significant differences from the unlensed one.    
\label{Inputwaveform}}
\end{figure*}

In practice, we simulate such a system with a scaling factor of $750$. As such, the lens is $10^5 M_{\odot}$ in our new simulations. We choose the radius of the cylinder as $R_{\rm cylinder}= 48.73 \rho_s$ and the length as $L_{\rm cylinder}= 146.18 \rho_s$, where $\rho_s$ is the horizon of the lens black hole. The simulation has a refinement of $2^8$ with a total DOF of $8.055\times 10^8$, the same as before.

Figure~\ref{Inputwaveform} shows the input waveform (the solid black line) and the temporal lensed waveform (the orange line), taken at $x=0.023333[{\rm Sec}]\approx 73\rho_s$ along the $x$-axis. This position is far away from the horizon, where the impact of the black hole on wave propagation is small. Indeed, as pointed out in~\cite{PhysRevD.103.064047}, since the amplitude of the lensed waveform completely degenerates with the luminosity distance, the detectability of the lensed waveform comes from the changes of shape relative to the unlensed one. We therefore  normalize the amplitude of the lensed waveform by its maximum value. The initial amplitude of the input waveform is then re-scaled to match the lensed waveform. Note that its time is also shifted for the purpose of comparison.   

Compared with the unlensed waveform, the lensed waveform has two important features after passing through the black hole. First, unlike in the flat spacetime, the shape of the lensed waveform can be permanently changed due to the effect of back-scattering in strong gravity. One obvious example for this effect is the wavelet with a sharp trailing wavefront as already shown in Fig.~\ref{powerlawtail}. After passing through the black hole, the wavelet has a clear tail. Second, the relative strength of amplitude after the lens is frequency dependent. This feature is significant even in the weak field limit~\cite{Takahashi_2003}.

As a result, because of the above two effects, the lensed waveform is much longer than the input waveform in the merger and ringdown phases as shown Fig.~\ref{Inputwaveform}. The relative strength of the lensed waveform in the merger and ringdown also exhibits significant differences from the input waveform. 

\begin{figure*}[!htb]
{\includegraphics[width=\linewidth]{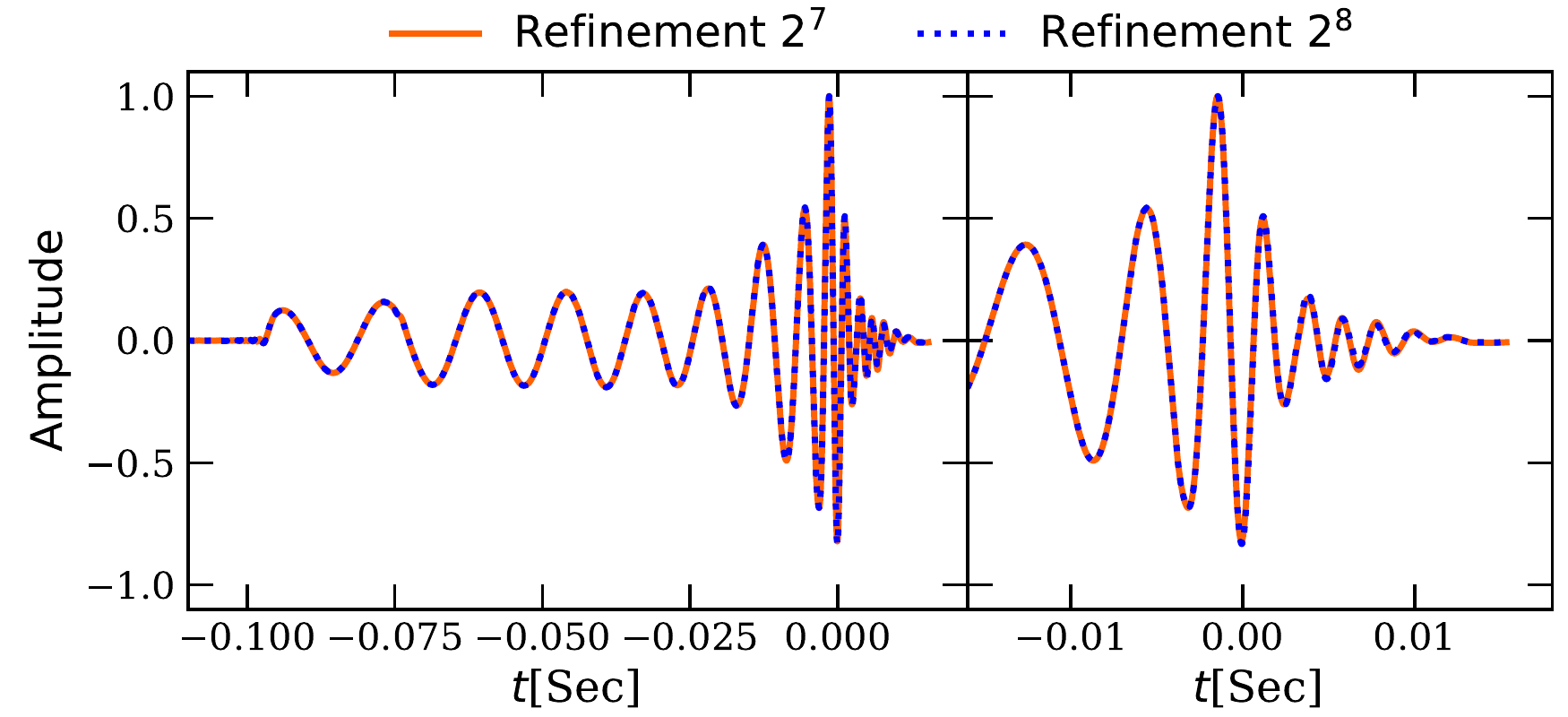}}
\caption{The numerical convergence test of the lensed waveform. The solid orange line shows the waveform from the simulation with a refinement of $2^7$ and the blue dashed line shows the waveform obtained from a higher resolution with a refinement of $2^8$. The left panel shows the lensed waveforms and the right panel shows an enlargement of the waveform for the merger and ringdown phases. Moreover, since the initial wavelength is much larger than the one used in our previous wavelet test, there are almost no wavefront shocks in the lensed waveform.   
\label{Convergency_test}}
\end{figure*}

Figure~\ref{Convergency_test} shows the numerical convergence test of the lensed waveforms. The solid orange line is the waveform from the simulation with a refinement of $2^7$ and the blue dashed line is the waveform obtained from a higher resolution with a refinement of $2^8$. The left panel shows the lensed waveform and the right panel shows an enlargement of the waveform for the merger and ringdown phases. The numerical results from both $2^7$ and $2^8$ refinements show a good agreement. Moreover, since the initial wavelength is much larger than the one used in our previous wavelet test, there are almost no wavefront shocks in the lensed waveform. 

\subsection{Hamiltonian constraint}
The evolution equation for the Hamiltonian constraint  is given by
\begin{align}
\frac{\partial \mathcal{H}}{\partial t} =-D_i(NM^i)+2N K\mathcal{H}-M^iD_iN\,.
\end{align}
In the background spacetime, the above equation is trivial since both $\mathcal{H}$ and $M^i$ vanish in the background $\mathcal{H}=0\,,M^i=0$. Taking the linear perturbation of the above equation, we obtain 
\begin{align}
\frac{\partial \delta \mathcal{H}}{\partial t} &=-2\delta M^i D_i N -2M^i D_i \delta N -\delta N D_i M^i - N \delta (D_i M^i) \nonumber \\
&=-2\delta M^i D_i N\nonumber\,.
\end{align}
If the perturbed constraints are satisfied at $t=0$, 
\begin{equation}
\left\{
\begin{aligned}
\delta \mathcal{H}|_{t=0}&=0\nonumber \\
\delta M^i|_{t=0}&=0
\end{aligned}
\right. \,,
\end{equation}
then $\delta \mathcal{H}=0$ and $\delta M^i=0$  are preserved by the wave equation Eq.~(\ref{perturbedwaveequation}).

In fact, by straightforward calculation, we obtain
\begin{align}
\delta \mathcal{H}=\frac{1}{2}\delta R = \frac{1}{2}\left(\delta \gamma^{ij} R_{ij} + \gamma^{ij} \delta R_{ij}\right)\,.
\end{align}
From Eq.~(\ref{wavetrace}), we find
\begin{align}
\gamma^{ij}\delta R_{ij}=-\frac{1}{2N^2}\frac{\partial^2h}{\partial t^2}-\gamma^{im}\gamma^{jn}h_{mn}\frac{D_iD_j N}{N}\,.
\end{align}
Therefore, we obtain
\begin{align}
\delta \mathcal{H}&=\frac{1}{2}\delta R \nonumber \\
&=-\frac{1}{4N^2}\frac{\partial^2h}{\partial t^2}-\frac{\gamma^{im}\gamma^{jn}h_{mn}}{2}\left(R_{ij} -\frac{D_iD_j N}{N}\right)\nonumber\\
&=-\frac{1}{4N^2}\frac{\partial^2h}{\partial t^2}\,.
\end{align}

From the above equation, the perturbed Hamiltonian constraint $\delta \mathcal{H} =0$ is satisfied as long as $h_{ij}$ is trace-free. Since $h_{ij}$ is taken to be exactly trace-free in our scheme, the perturbed Hamiltonian constraint holds exactly in our approach.

\section{conclusions\label{sec::conclusions}}
In this paper, we have substantially extended our previous work~\cite{He:2021hhl} simulating the propagation of GWs in a potential well in the weak field limit to the regime of strong gravity. We have developed a code that is capable of studying the propagation of GWs in the spacetime of a Schwarzschild black hole.
Based on the 3+1 form of Einstein's equation, we calculate the perturbation equations up to linear order in a self-consistent manner. We have shown explicitly that these equations are covariant under the arbitrary infinitesimal coordinate transformation.
Moreover, unlike the conventional perturbed equations in a black hole~\cite{PhysRev.108.1063,PhysRevLett.24.737,PhysRevD.2.2141,Teukolsky}, our formalism is less coordinate dependent. There are no coordinate singularities in our approach. As such, no regularity conditions are needed in our analyses.

To numerically solve these equations, we firstly derived their {\it weak formulation}. Then we adopted the cGFEM, based on the publicly available code DEAL.II. We evolve the perturbed equations in a maximal slicing $\delta N =0$. Compared with the scheme presented in ~\cite{PhysRevD.70.104007,PhysRevD.77.084007} in numerical relativity, a notable feature of our work is that the wave speed is no longer a constant but varies in space, which equals the speed of light observed by an asymptotic observer. Since the wave speed vanishes at the horizon, a particular advantage of our approach is that it can naturally avoid boundary conditions at the horizon. 

Based on our code, we have first simulated a finite wave train of GWs passing through the spacetime of a Schwarzschild black hole. Since the leading wavefront represents the transfer of energy from one place to another, it is subject to the constraint of causality. Moreover, since the speed of the leading wavefront equals the speed of light rays, the leading wavefront is a null hypersurface and its integral curves are null geodesics. As a result, the leading wavefront can be traced by null geodesics, which provides a robust way to test our numerical results. We find that our numerical simulations agree with the theoretical predictions very well.

Besides the leading wavefront, we have also studied the evolution of wave zones of GWs in the spacetime of the Schwarzschild black hole. Behind the black hole, the wave zone is wildly twisted, which has a complicated geometry. Moreover, we find that the back-scattering due to the interaction between GWs and the background curvature is strongly dependent on the direction of the propagation of the trailing wavefront relative to the black hole. For waves that are far away from the horizon, the trailing wavefront travels nearly perpendicular to the radius. There is no back-scattering effect and the trailing wavefront is still clear. However, in regions that are around the $x$-axis, where the trailing wavefront travels along the radius of the black hole, we find that a clear tail behind the trailing wavefront emerges. However, such a tail does not appear before the black hole as in this case the trailing wavefront travel faster than the leading wavefront. 

Moreover, we have also simulated a continuous wave train passing through the black hole. We find that, unlike geometric optics, GW signals cannot be sheltered by the black hole due to the wave nature of GWs. GWs do not form a caustic behind the black hole at scales that are comparable to their wavelength. Instead, a strong beam and an interference pattern appear in the overlaps of the wave zones behind the black hole along the optical axis. The wave functions in our simulations are well defined on the optical axis, which are different from those in the scattering theory, where the wave functions are divergent along the optical axis behind the black hole~\cite{PhysRevD.52.1808,PhysRevD.18.1798}. The reason for such differences is due to the boundary conditions imposed in the scattering theory, which implicitly assume that the scattered waves are spherical. This, however, is not the case in our simulations. The wavefront of the outgoing GWs in our simulations has a complicated geometric shape.

For a realistic input waveform generated by binary black holes, we find that when passing through the lens back hole, the lensed waveform in the merger and ringdown phases is much longer than that of the input waveform due to the effect of back-scattering in strong gravity. The relative strength of the lensed waveform in the merger and ringdown phases also exhibits significant differences from that of the input waveform.

Further, it is worth noting that due to the linearity of the wave equation and the geometric unit, the parameters used in our code and numerical simulations can be rescaled to other values that are of astrophysical interest. For instance, if we want to obtain the results for a black hole with a mass of $M=10 M_{\odot}$, we only need to rescale the temporal and spatial axes of our numerical results by $1/3 \times 10^{-4}$ times. 

Finally, this paper mainly focuses on the numerical technique aspect. In a companion paper, we will present a comprehensive analysis of the detectability of GWs passing through an isolated Schwarzschild black hole against the sensitivities of current and future GW detectors using Bayesian Inference~\cite{DelPozzo:2014cla}. Moreover, since our formalism does not involve artificial boundary conditions, our work can be extended to study the physical origin of exciting QNMs in the spacetime of a black hole. Our formalism can also provide a potential way to study GWs produced by the generic extreme mass ratio inspirals (EMRIs) in the time domain  ~\cite{PhysRevD.61.084004,PhysRevD.73.024027}. It is also of great interest to extend our work to the spacetime of a Kerr black hole~\cite{Baraldo:1999ny}. Detailed analyses of these issues will be presented in a series of follow-up papers in the future. 

\vspace{5mm}

\noindent {\bf Acknowledgments}
J.H.H. acknowledges support of Nanjing University. This work is supported by the National Key R$\&$D Program of China (Grant No. 2021YFC2203002), the National Natural Science Foundation of China (Grants No. 12075116, No.12150011), the science research grants from the China Manned Space Project (Grant No.CMS-CSST-2021-A03). The Natural Science Foundation of the Jiangsu Higher Education Institutions of China (Grant No. 22KJB630006). The numerical calculations in this paper have been done on the computing facilities in the High Performance Computing Center (HPCC) of Nanjing University.

\bibliography{myref}

\section*{Appendix}
\subsection{general covariance of the perturbation equations\label{covariantproof}}
In this appendix, we show the general covariance of the perturbation equations. We start with the wave equation
\begin{align}
\frac{\partial^2}{\partial t^2}h_{ij}&= 2N\delta(D_iD_jN)-2N^2\delta R_{ij}-2N\delta N R_{ij}\nonumber\\
&=2N[D_iD_j\delta N-\frac{1}{2}\gamma^{kl}\partial_kN(D_i h_{lj}+D_j h_{il}-D_l h_{ij})\nonumber\\
&-N\delta R_{ij}-\delta N R_{ij}]\,.\label{secondwaveper}
\end{align}
Inserting the following infinitesimal coordinate transformation into the RHS of the above equation
\begin{equation}
\left\{
\begin{aligned}
\delta \tilde{N}&\rightarrow \delta N - \eta^kD_k N\\
\tilde{h}_{ij}&\rightarrow h_{ij} - D_j\eta_i-D_i\eta_j \\
\delta \tilde{R}_{ij}&\rightarrow \delta R_{ij} - R_{ik}D_j\eta^k-R_{kj}D_i\eta^k - \eta^kD_k R_{ij} \label{infinitrans2}
\end{aligned} \right.\,,
\end{equation}
we obtain
\begin{align}
\frac{\widetilde{{\rm RHS}}}{2N}&\rightarrow \frac{{\rm RHS}}{2N}-D_iD_j(\eta^kD_kN)\nonumber\\
&+\frac{1}{2}D^lN(D_iD_l\eta_j+D_iD_j\eta_l+D_jD_i\eta_l\nonumber \\
&+D_jD_l\eta_i-D_lD_j\eta_i-D_lD_i\eta_i)\nonumber\\
&+N(R_{ik}D_j\eta^k+R_{kj}D_i\eta^k+\eta^kD_kR_{ij})\nonumber\\
&+\eta^kD_kR_{ij}\,.\label{wavetrans}
\end{align}
From the definition of the Riemann tensor, we have
\begin{align}
D_iD_l\eta_j&=D_lD_i\eta_j-R_{ilmj}\eta^m\nonumber\\
D_jD_l\eta_i&=D_lD_j\eta_i-R_{jlmi}\eta^m\nonumber\\
D_jD_i\eta_l&=D_iD_j\eta_l-R_{jiml}\eta^m\,.
\end{align}
We expand the first term as
\begin{align}
D_iD_j(\eta^kD_kN)&=D_kND_iD_j\eta^k+D_j\eta^kD_iD_kN\nonumber\\
&+D_i\eta^kD_jD_kN+\eta^kD_iD_jD_kN\,.
\end{align}
Further noting that $D_iD_jN=NR_{ij}$, from the above expressions,
Eq.~(\ref{wavetrans}) reduces to
\begin{align}
\frac{\widetilde{{\rm RHS}}}{2N}&\rightarrow \frac{{\rm RHS}}{2N}- \eta^kD_iD_jD_kN+\eta^kD_k(NR_{ij})\nonumber\\
&-\frac{1}{2}D^lN(R_{jiml}+R_{ilmj}+R_{jlmi})\eta^m
\,.
\end{align}
The terms with Riemann tensor can be further simplified using the circling identity of Riemann tensor $$R_{jiml}+R_{mjil}+R_{imjl}=0\,,$$ which gives
\begin{align}
\frac{1}{2}D^lN(R_{jiml}+R_{ilmj}+R_{jlmi})\eta^m=D^lNR_{jlki}\eta^k\,.
\end{align}
Further noting that 
\begin{align}
\eta^kD_iD_jD_kN&=\eta^kD_iD_kD_jN\nonumber\\
&=\eta^k(D_kD_iD_jN-R_{iklj}D^lN)\,,
\end{align}
we find
\begin{align}
\frac{\widetilde{{\rm RHS}}}{2N}&\rightarrow \frac{{\rm RHS}}{2N}- \eta^kD_iD_jD_kN+\eta^kD_k(NR_{ij})\nonumber\\
&-\frac{1}{2}D^lN(R_{jiml}+R_{ilmj}+R_{jlmi})\eta^m\nonumber\\
&=\frac{{\rm RHS}}{2N}-\eta^kD_kD_iD_jN+\eta^kD_k(NR_{ij})\nonumber\\
&=\frac{{\rm RHS}}{2N}
\,.
\end{align}
The above result shows that the RHS of Eq.~(\ref{secondwaveper}) does not change its format under arbitrary infinitesimal coordinate transformations. Since $- D_j\eta_i-D_i\eta_j$ is time independent, the LHS of Eq.~(\ref{secondwaveper}) $\frac{\partial^2}{\partial t^2}h_{ij}$ does not change its format as well. Equation~(\ref{secondwaveper}), therefore, is covariant.

Not only is Eq.~(\ref{secondwaveper}) covariant, we can also show that
\begin{align}
\delta R_{ij}&=\frac{1}{2}(D^lD_i h_{lj}+D^lD_jh_{il}-D^lD_lh_{ij})\nonumber\\
&-\frac{1}{2}D_jD_i(\gamma^{kl}h_{lk})\,,
\end{align}
is covariant.
To do this, we insert the infinitesimal coordinate transformation of $h_{ij}$ into the RHS of the above equation
\begin{align}
\tilde{h}_{ij}&\rightarrow h_{ij} - \gamma_{ik}D_j\eta^k-\gamma_{kj}D_i\eta^k\,,
\end{align}
we obtain
\begin{align}
\widetilde{{\rm RHS}}\rightarrow &{\rm RHS}-\frac{1}{2}D^lD_iD_l\eta_j-\frac{1}{2}D^lD_iD_j\eta_l-\frac{1}{2}D^lD_jD_l\eta_i\nonumber\\
&-\frac{1}{2}D^lD_jD_i\eta_l+\frac{1}{2}D^lD_lD_i\eta_j+\frac{1}{2}D^lD_lD_j\eta_i\nonumber\\
&+D_iD_jD^k\eta_k\,.
\end{align}
From the definition of the Riemann tensor, we have
\begin{align}
D^lD_lD_i\eta_j&=D^lD_iD_l\eta_j-\eta^kD^lR_{likj}-D^l\eta^k R_{likj}\nonumber\\
D^lD_lD_j\eta_i&=D^lD_jD_l\eta_i-\eta^kD^lR_{ljki}-D^l\eta^k R_{ljki}\,.
\end{align}
Further note that
\begin{align}
D_iD_jD_k\eta_k&=D_iD_kD_j\eta^k-D_i(R_{jm}\eta^m)\nonumber\\
&=D_kD_iD_j\eta^k-R_{ikmj}D^m\eta^k-R_{im}D_j\eta^m\nonumber\\
&-\eta^mD_iR_{jm}-R_{jm}D_i\eta^m\,
\end{align}
and
\begin{align}
D^lD_jD_i\eta_l&=D^lD_iD_j\eta_l-D^l(R_{jiml}\eta^m)\,,
\end{align}
we obtain
\begin{align}
\widetilde{{\rm RHS}}\rightarrow &{\rm RHS}+\frac{1}{2}D^l\eta^k(R_{jikl}+R_{ikjl}+R_{kjil})\nonumber\\
&-R_{ik}D_j\eta^k-R_{jk}D_i\eta^k-\eta^kD_iR_{jk}\nonumber\\
&+\frac{1}{2}\eta^k(D^lR_{jikl}-D^lR_{likj}-D^lR_{ljki})\,.
\end{align}
Using the circling identity of Riemann tensor $R_{jikl}+R_{ikjl}+R_{kjil}=0$
and the Bianchi identity
\begin{align}
D^lR_{jikl}&=D_iR_{jk}-D_jR_{ik}\nonumber\\
D^lR_{likj}&=-D_jR_{ki}+D_kR_{ji}\nonumber\\
D^lR_{ljki}&=-D_iR_{kj}+D_kR_{ij}\,,
\end{align}
we find that
\begin{align}
\widetilde{{\rm RHS}}\rightarrow &{\rm RHS}-R_{ik}D_j\eta^k-R_{jk}D_i\eta^k-\eta^kD_kR_{ij}\,,
\end{align}
which is consistent with doing the infinitesimal coordinate transformation of $\delta R_{ij}$ directly. The expression of $\delta R_{ij}$ in terms of $h_{ij}$, therefore, is covariant.

Next, we show that the perturbed conservation laws are covariant as well
\begin{align}
D^l \Gamma_l-D^lD_l h=\gamma^{im}\gamma^{jn} h_{mn}R_{ij}\,.\label{Dgammaapp}
\end{align}
Under the infinitesimal coordinate transformation Eq.~(\ref{infinitrans2}), $\Gamma_l$ and $h$ in Eq.~(\ref{Dgammaapp}) transform as
\begin{align}
&D^l\tilde{\Gamma}_l\rightarrow D^l\Gamma_l - D_lD_mD^l\eta^m - D_lD_mD^m\eta^l\nonumber \\
&D^lD_l \tilde{h} \rightarrow D^lD_l h - 2 D^lD_l D_m\eta^m \,.
\end{align}
From the definition of Ricci tensor, we obtain
\begin{align}
(D_lD_m-D_mD_l)\eta^m=-R_{lk}\eta^k\,.
\end{align}
Taking the derivative $D^l$ of above equation, we find
\begin{align}
D^lD_l D_m\eta^m=D^lD_m D_l\eta^m-\eta^mD^lR_{lm}-R_{lm}D^l\eta^m\,.
\end{align}
Since the background scalar curvature vanishes $R=0$, from the Bianchi identity we have $2D^lR_{lm}=D_m R =0$.
Then it follows that
\begin{align}
D^lD_l D_m\eta^m=D^lD_m D_l\eta^m-R_{lm}D^l\eta^m\,.
\end{align}
The infinitesimal coordinate transformation on the left hand side of Eq.~(\ref{Dgammaapp}) gives
\begin{align}
\widetilde{{\rm RHS}} \rightarrow & {\rm RHS} - D_lD_mD^l\eta^m - D_lD_mD^m\eta^l + 2 D^lD_l D_m\eta^m\nonumber\\
=&D_lD_mD^l\eta^m - D_lD_mD^m\eta^l-2 R_{lm}D^l\eta^m\,.
\end{align}
Further note that
\begin{align}
D_lD_mD^l\eta^m &= D_mD_lD^l\eta^m - \tensor{R}{_l_m_k^l}D^k\eta^m- \tensor{R}{_l_m_k^m}D^l\eta^k\nonumber\\
&=D_mD_lD^l\eta^m + R_{mk}D^k\eta^m-R_{lk}D^l\eta^k\nonumber\\
&=D_mD_lD^l\eta^m\nonumber\\
&=D_lD_mD^m\eta^l\,,
\end{align}
we obtain
\begin{align}
D^l \tilde{\Gamma}_l-D^lD_l 
\tilde{h}&=D^l \Gamma_l-D^lD_l h-2 R_{lm}D^l\eta^m\nonumber\\
         &=\gamma^{im}\gamma^{jn} h_{mn}R_{ij}-2 R_{ij}D^i\eta^j\nonumber\\
	 &=\gamma^{im}\gamma^{jn} \tilde{h}_{mn}R_{ij}\,.
\end{align}
The above equation demonstrates that Eq.~(\ref{Dgamma}) does not change its format under arbitrary infinitesimal coordinate transformation, which means that Eq.~(\ref{Dgamma}) is covariant. 
\subsection{Null geodesics\label{Nullgeodesic}}
The trajectories of null geodesics in the spacetime of a Schwarzschild black hole are given by
\begin{align}
\frac{d^2\mu}{d\phi^2}+\mu=3M\mu^2\,,
\end{align}
where $r$ and $\phi$ are the radius and azimuthal angle in the Schwarzschild coordinates. $\mu$ is the inverse of the radius $\mu=1/r$. We then make coordinate transformations  
\begin{equation}
\left \{
\begin{aligned}
r&=\frac{(M+2\rho)^2}{4\rho}\\
\rho&=\frac{1}{2}\left(r-M+\sqrt{r^2-2Mr}\right)
\end{aligned}
\right. \,,\label{rho2r}
\end{equation}
where $\rho$ is the radius in isotropic coordinates. With the notion $\mu'=1/\rho$, the geodesic equations in the isotropic coordinates are given by 
\begin{equation}
\left \{
\begin{aligned}
\frac{d^2\mu'}{d\phi^2}&=\frac{3M-r}{\rho^2}\frac{d\rho}{dr}\\
\frac{d\rho}{dr}&=\frac{1}{2}\left(1+\frac{r-M}{\sqrt{r^2-2Mr}}\right)
\end{aligned}
\right. \,.
\end{equation}
In the above equations, $r$ can be replaced by $\rho=1/\mu'$ from Eq.~(\ref{rho2r}). After obtaining the geodesics, the length of trajectories can be obtained by
\begin{align}
dl=\frac{1}{\mu'}\sqrt{\frac{1}{{\mu'}^2}\left(\frac{d\mu'}{d\phi}\right)^2+1}d\phi\,.
\end{align}
Then the total asymptotic time along the null geodesics can be computed by  
\begin{align}
t=\int \frac{dl}{c}\,,
\end{align}
where $c$ is the isotropic speed of wave defined in Eq.~(\ref{definationspeed}). 
\end{document}